\documentclass{ieeeaccess}
\DeclareFontShape{T1}{formata}{m}{sl}{<-> ssub * formata/m/it}{}
\pdfmapfile{=t1-formata.map}

\usepackage{cite}
\usepackage{amsmath,amssymb,amsfonts}
\usepackage{algorithmic}
\usepackage{graphicx}
\usepackage{textcomp}
\usepackage{float}
\usepackage{bm}
\makeatletter
\AtBeginDocument{\DeclareMathVersion{bold}
\SetSymbolFont{operators}{bold}{T1}{times}{b}{n}
\SetSymbolFont{NewLetters}{bold}{T1}{times}{b}{it}
\SetMathAlphabet{\mathrm}{bold}{T1}{times}{b}{n}
\SetMathAlphabet{\mathit}{bold}{T1}{times}{b}{it}
\SetMathAlphabet{\mathbf}{bold}{T1}{times}{b}{n}
\SetMathAlphabet{\mathtt}{bold}{OT1}{pcr}{b}{n}
\SetSymbolFont{symbols}{bold}{OMS}{cmsy}{b}{n}
\renewcommand\boldmath{\@nomath\boldmath\mathversion{bold}}}
\makeatother

\usepackage[T1]{fontenc}
\usepackage{textcomp}
\usepackage{fancyhdr}
\fancypagestyle{plain}{
  \fancyhf{}
  
  \fancyhead[C]{\scriptsize  This article has been published in \textit{IEEE Access}.}
}
\usepackage{amsfonts}
\usepackage{array}
\newcommand{\orcidicon}[1]{\href{https://orcid.org/#1}{\raisebox{-0.2ex}{\includegraphics[height=1.6ex]{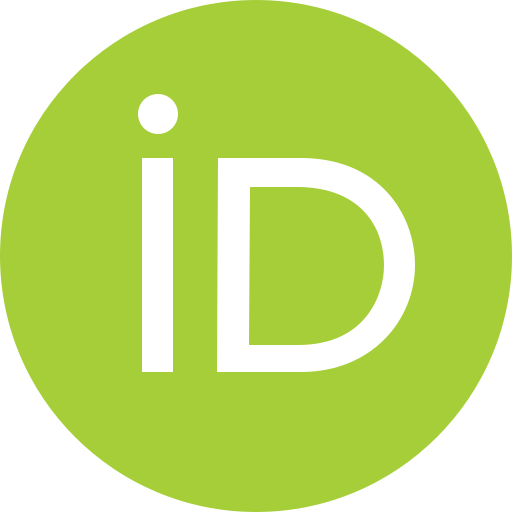}}}}
\usepackage{textcomp}
\usepackage{stfloats}
\usepackage{url}
\usepackage{soul}\usepackage{threeparttable}
\usepackage{algorithmic}
\usepackage[colorlinks=false, linkcolor=black, citecolor=black, urlcolor=black, pdfborder={0 0 0}, unicode]{hyperref}    
\makeatother

\definecolor{darkgreen}{rgb}{0.0, 0.5, 0.0}
\definecolor{darkcyan}{rgb}{0.0, 0.35, 0.55}

\def\BibTeX{{\rm B\kern-.05em{\sc i\kern-.025em b}\kern-.08em
    T\kern-.1667em\lower.7ex\hbox{E}\kern-.125emX}}

\begin{document}
\history{}
\doi{\href{https://doi.org/10.1109/ACCESS.2025.3642613}{10.1109/ACCESS.2025.3642613}}

\title{Low-Complexity Frequency-Dependent Linearizers Based on Parallel Bias-Modulus and Bias-ReLU Operations}
\author{\uppercase{Deijany Rodriguez Linares}\orcidicon{0009-0004-1846-9496}, \IEEEmembership{Graduate Student Member, IEEE}, and
\uppercase{H\AA kan Johansson}\orcidicon{0000-0001-6329-9132}, \IEEEmembership{Senior Member, IEEE}}

\address{Department of Electrical Engineering, Linköping University, 58183 Linköping, Sweden}

\tfootnote{This work belongs to the project "Baseband Processing for Beyond 5G Wireless", which is financially supported by ELLIIT.}

\markboth
{D. Rodriguez-Linares and H. Johansson: Low-Complexity Frequency-Dependent Linearizers}
{D. Rodriguez-Linares and H. Johansson: Low-Complexity Frequency-Dependent Linearizers}

\corresp{Corresponding author: Deijany Rodriguez Linares (e-mail: deijany.rodriguez.linares@liu.se).}

\begin{abstract}
    This paper introduces low-complexity frequency-dependent (memory) linearizers designed to suppress nonlinear distortion in analog-to-digital interfaces. Two different linearizers are considered, based on nonlinearity models which correspond to sampling before and after the nonlinearity operations, respectively. The proposed linearizers are inspired by convolutional neural networks but have an \mbox{order-of-magnitude} lower implementation complexity compared to existing neural-network-based linearizer schemes. The proposed linearizers can also outperform the traditional parallel Hammerstein linearizers even when the nonlinearities have been generated through a Hammerstein model. Further, a design procedure is proposed in which the linearizer parameters are obtained through matrix inversion. This eliminates the need for costly and time-consuming iterative nonconvex optimization that is traditionally associated with neural network training. The design effectively handles a wide range of wideband multi-tone signals and filtered white noise. Examples demonstrate significant \mbox{signal-to-noise-and-distortion} ratio (SNDR) improvements of about $20$--$30$ dB, as well as a lower implementation complexity than the Hammerstein linearizers.
\end{abstract}


\titlepgskip=-21pt

\maketitle
\thispagestyle{plain} 
\section{Introduction}\label{sec:Introduction}

Conversions between analog and digital signals constitute fundamental functions that will always be required, as the physical world is analog by nature whereas the signal processing is primarily carried out in the digital domain. These fundamental functions are implemented using analog-to-digital converters (ADCs) and digital-to-analog converters (DACs) as well as other components required before/after the ADC/DAC, like filters, power amplifiers, and mixers. The overall converters are referred to as analog-to-digital and digital-to-analog interfaces (ADIs and DAIs, respectively). With the ever-increasing demands for high-performance and low-cost signal processing and communication systems, there is a need to develop ADIs and DAIs with higher performance in terms of data rates (related to bandwidths), effective resolution {[}effective number of bits (ENOB) determined by the signal-to-noise-and-distortion ratio (SNDR)\footnote{For a full-scale sinusoidal, $\text{ENOB}=(\text{SNDR}-1.76)/6.02$.}{]}, and low implementation cost (small chip area and low energy consumption) \cite{Bjornson24, Tsinos25, haider22, Chu24, Zhao25}. This paper focuses on ADIs.

In analog circuits, the ENOB is degraded by nonlinearities and linearization techniques are therefore required to reach a large ENOB which is required in many applications. For instance, decoding high-order modulation schemes such as 1024–quadrature amplitude modulation (1024-QAM) requires SNRs around 35\,dB~\cite{qam_snr_ref}. High SNDR is also needed for recovering weak signals that may otherwise be masked by stronger ones, due to analog front-end nonlinearities and I/Q imbalance, where over 50\,dB distortion suppression is typically required \cite{valkama06,valkama01}.

Linearization can however also be utilized for low/medium ENOBs. For example, it enables the use of one- or few-bit ADCs, with performance below the targeted ENOB which is subsequently achieved through digital processing. Thereby the energy consumption of the ADCs can be reduced substantially as their energy consumption is very high for large ENOBs, especially when the data rate is also high \cite{Murmann_2021a}. For example, the 12-bit ADC12DJ5200SE from Texas Instruments \cite{TI_ADC12DJ5200SE}, dissipates approximately 4 Watts at a 10.4 GS/s data rate in single-channel mode. It is particularly important to reduce the ADC power in applications requiring many ADCs like in massive-MIMO communication systems \cite{Marzetta_2016,Bjornson24}. However, to reach the targeted ENOB while keeping the overall energy consumption low, it is vital to develop energy-efficient digital linearizers\footnote{There are trade-offs between analog and digital implementation complexities for a targeted resolution (SNDR performance), but such trade-off studies are beyond the scope of this paper. The objective here is to, given nonlinearities emanating from analog errors, reduce the complexity of the digital linearization. To this end, the paper introduces novel linearizer structures and demonstrates through extensive simulations that they offer lower computational complexity than the benchmark Hammerstein linearizers for the same SNDR performance.}.

The nonlinear errors can be suppressed by utilizing an a-priori assumed system model and parameter/order identification within the model. One can in principle use a Volterra series and its inverse \cite{Frank1996SamplingRF}, but the identification and compensation may become very complex and prohibitive. Alternatively, aiming at a reduced complexity, one may use specific structural system models \cite{Chen_95}, in particular parallel linear-nonlinear-linear (LNL) structures which \mbox{include} polynomial-based parallel Hammerstein, Wiener, and Wiener-Hammerstein structures. However, also for these options, there is a need to find an appropriate system model for each new circuit and their complexity still tends to be high. As an attempt to alleviate these problems, the use of neural-network based linearizers have appeared as an alternative approach, as they can model high-dimensional problems without explicit analytical expressions. However, the neural-network schemes that have appeared so far in this context typically have large design and implementation complexities. The ADI linearization papers that exist have reported schemes for which several hundreds or even thousands of multiplications are required per corrected sample \cite{Xu2019DeepLearningADC,Chen20,DENG_202063,Peng_2021,Chen21,Fayazi21,Zhai22,Zhifei24,Peng_2024}. {For the Hammerstein linearizers (used as the benchmark in this paper, see below) and proposed linearizers, the number of multiplications required is about an order of magnitude lower, as the examples in the paper will demonstrate (see Section~\ref{subsec:complexity1}). Further, compared with the Hammerstein linearizers, the proposed ones offer savings between a few percent up to about 60 percent depending on the scenario (see Sections~\ref{subsec:complexity1} and~\ref{subsec:complexity2}).} 

\subsection{Contribution and Relation to Previous Works}\label{subsec:Contribution}

Traditional linearizers based on parallel LNL systems utilize linear filters and polynomial nonlinearity terms $v^{p}(n)$, for a set of $p$-values, where $v(n)$ is the distorted signal. Common polynomial linearizers are the parallel Hammerstein system\footnote{Throughout the paper, parallel Hammerstein will for simplicity be referred to as Hammerstein.} (seen in Fig. \ref{Flo:Hammerstein-scheme} in Section \ref{sec:Signal-model}), and the corresponding Wiener system which is obtained by interchanging the nonlinearities and filters \cite{Chen_95}. {Both systems are widely used, and they have the same complexity for the same number of branches and filter order. They are generally not equivalent though (except for the memoryless case), and the best choice depends on the system at hand that should be linearized \cite[Chapt.~5]{ljung1999system}. However, this concerns finding appropriate nonlinear models, which is beyond the scope of this paper. Instead, the focus here is to introduce a novel linearizer (explained below) and show that it can outperform classical polynomial-based linearizers. To this end, we have chosen the Hammerstein linearizer as the benchmark, as it is more straightforward to design, and has been shown to be appropriate for many practical circuits and systems \cite{Mao19, Gilabert2005, Sadeghpour2011, haider22}. In particular, it will be demonstrated that the proposed linearizer can outperform the Hammerstein linearizer even when the nonlinearities have been generated through a Hammerstein model.}

In the proposed novel linearizer (seen in Fig. \ref{Flo:proposed-scheme1} in Section \ref{sec:Proposed-Linearizer}), the nonlinearity terms $v^{p}(n)$ in the Hammerstein linearizer are replaced by the simpler nonlinearities $|v(n)|$ or $\max\{0,v(n)\}$ {[}rectified linear unit (ReLU){]} as they can be implemented in hardware with low complexity \cite{Tarver_2019, deiro23}. In addition, bias values are added before the nonlinearities. The proposed linearizer is inspired by convolutional neural networks, but the neural-network schemes that have appeared earlier in the literature in this context have a very high implementation complexity as mentioned earlier. The implementation complexity of the proposed linearizer is an order-of-magnitude lower, primarily because it does not contain the multiple layers and interconnections used in traditional neural networks. As mentioned above, it will also be demonstrated that it can outperform the Hammerstein linearizer. Further, a design procedure is proposed in which the parameters (filter coefficients) are obtained through matrix inversion. Thereby, one can eliminate the costly and time-consuming iterative nonconvex optimization that is traditionally used when training neural networks. Moreover, the design and evaluation incorporate a large set of wideband multi-tone signals and filtered white noise. In the previous works \cite{Xu2019DeepLearningADC,Chen20,DENG_202063,Peng_2021,Chen21,Zhai22,Peng_2024}, the evaluations have only included single-tone and few-tone signals or narrowband multi-tone signals. Our simulations show SNDR improvements up to about $20$--$30$ dB for a wide range of wideband signals covering most of the Nyquist band. {It is noted here that the proposed linearizers are not limited to low- or high-resolution ADCs, but to investigate their distortion-suppression capabilities, the focus in the examples is on higher resolutions (10-12 bits).}

A special case of the proposed linearizer is obtained when the filters are replaced with single parameters, i.e., 0th-order filters. This corresponds to the memoryless linearizer which we introduced in \cite{deiro23} and was shown to outperform the Hammerstein linearizer. The use of a memoryless linearizer is typically sufficient for narrow to medium bandwidths and resolutions. To reach higher resolutions over wider frequency bands, one needs to incorporate memory (filters) in the modeling and linearization which is in focus here. Hence, this paper extends the memoryless linearizer in \cite{deiro23} to memory
(frequency dependent) linearizers by incorporating arbitrary-order filters.

Moreover, it is often assumed that the ADI nonlinearity distortion can be modeled as occurring after sampling. If the nonlinearity distortion is incurred before sampling, problems arise though due to undersampled nonlinearities when the desired (original) signal covers the whole (or most of) the Nyquist band. In the frequency-independent (memoryless) case, the two models are equivalent as sampling and static nonlinearities commute. However, in general, the two models are not equivalent. As shown in \cite{Tsimbinos_98}, it is still possible though to recover the desired signal from the undersampled distorted signal, as long as the desired signal is Nyquist sampled. In practice, this can be carried out by incorporating interpolation in the linearization \cite{Vansebrouck_2017}. This paper considers this case as well, by extending the memoryless linearizer in \cite{deiro23} to incorporate both filters and additional interpolation. To this end, discrete-time equivalences of the linearizer schemes are derived which were not considered in \cite{Vansebrouck_2017}. 

It is also noted here that neural networks have been explored for predistorting power amplifiers (PAs) in DAIs \cite{Tarver_2019,Liu_2022,Li23,Jiang23, Ghazanfarianpoor2023, Prasad_2023, Prasad_2024, Hosseini2024, haider22}, but they are not directly transferable to ADI linearization. A significant difference is that the predistorted signal must be oversampled to enable distortion cancellation in the analog domain, which is not required for ADI linearization as mentioned above. A second major difference is that in ADIs, real-valued signals are typically considered, and the aim is to suppress all nonlinearity terms. For predistortion techniques in DAIs, one normally assumes a complex (I/Q) baseband signal in the digital domain, and it typically suffices to consider only odd-order terms in a polynomial model and to suppress the distortion centered around the signal band. The remaining distortion can in that case be removed by analog filtering.

\subsection{Outline} 
Following this introduction, the pre-sampling distortion model (sampling before incurred distortion) and the parallel Hammerstein linearizer are reviewed in Section \ref{sec:Signal-model}. Section \ref{sec:Proposed-Linearizer} introduces the proposed linearizer and design procedure for the pre-sampling distortion model, and also provides several simulation results. Then, Section \ref{sec:analog} considers the extension incorporating interpolation for the post-sampling distortion model (sampling after incurred distortion), including several simulations. Section \ref{sec:realdata} provides results for linearization of circuit-simulated data. Section \ref{sec:alternative-nonlinearities} discusses alternative nonlinear functions, whereas Section \ref{sec:conclusion} concludes the paper.
{
\begin{samepage}
\subsection*{Definitions, Notations, and Acronyms}
The main definitions, notations, and acronyms used throughout the paper are summarized in Table~\ref{tab:notations}.
\begin{table}[h!]
	\centering
	\caption{Definitions, Notations, and Acronyms}
	\label{tab:notations}
	\renewcommand{\arraystretch}{1.2}
	\begin{tabular}{p{0.25\linewidth} p{0.68\linewidth}}
	\hline
	\hspace{-3 pt}\textbf{Symbol / Acronym} & \textbf{Description} \\
	\hline
	$\mathbf{A}$ & Matrix used in the proposed design. \\		
	$a_p(k)$, $g_p(k)$ & Impulse responses of the nonlinear-distortion filters in the pre-sampling and post-sampling models. \\
	$\mathbf{b}$ & Vector used in the proposed design. \\
	$b_m$, $w_m(l)$ & Bias values and filter coefficients (weights) in branch $m$ of the proposed linearizer. \\
	$d_p(l)$ & Filter coefficients (weights) in branch $p$ of the Hammerstein linearizer. \\
	$\phi\left\{(\cdot)^k\right\}$ & Minimum number of multiplications required to generate the nonlinearities $(\cdot)^k$. \\
	$h_k(n)$ & Impulse responses of interpolation filters. \\
	$\mathrm{ReLU}(\cdot)$, $|\cdot|$ & Nonlinear operations in the proposed linearizer. \\
	$S(P_{K+1})$ & Number of multiplications required to create all the static nonlinearities in the Hammerstein post-sampling linearizer. \\
	$v^p(\cdot)$ & Nonlinear operations in the Hammerstein linearizer. \\
	$\mathbf{w}$ & Vector containing all filter coefficients (weights) of the proposed linearizer. \\
	$x_a(t)$ & Continuous-time signal. \\
	$x(n)$, $v(n)$, $y(n)$ & Desired, distorted, and compensated discrete-time signals. \\
	$\lambda$ & Parameter used in the $L_2$-regularization. \\
	ADC & Analog-to-digital converter. \\
	ADI & Analog-to-digital interface. \\
	DAC & Digital-to-analog converter. \\
	DAI & Digital-to-analog interface. \\
	dBFS & Decibel full scale. \\
	ENOB & Effective number of bits. \\
	ELU & Exponential linear unit. \\
	I/Q & In-phase/quadrature. \\
	LNL & Linear–nonlinear–linear. \\
	MIMO & Multiple-input multiple-output. \\
	QAM & Quadrature amplitude modulation.\\
	QPSK & Quadrature phase shift keying.\\
	SFDR & Spurious-free dynamic range. \\
	SNDR & Signal-to-noise-and-distortion ratio. \\
	SNR & Signal-to-noise ratio. \\
	\hline 
	\end{tabular}
\end{table}
\end{samepage}
}
\section{Pre-Sampling Distortion Model and Linearization}
\label{sec:Signal-model}
Consider a desired discrete-time signal (sequence) {$x(n)=x_{a}(nT)$}, which is a sampled version of a continuous-time signal $x_{a}(t)$ with a uniform sampling interval $T$. It is assumed that the Fourier transform $X_{a}(j\omega)$ is zero for $|\omega|>\omega_{c}$, with $\omega_{c}<\pi/T$, implying that $x_{a}(t)$ is bandlimited and satisfy the Nyquist criterion for uniform sampling of $x_{a}(t)$ with a sampling frequency of $1/T$ without introducing aliasing. In practice, the output of an ADC will not be $x(n)$ but a distorted version of it, say $v(n)$.

\begin{figure}[t]

\begin{centering}
\includegraphics[scale=0.58]{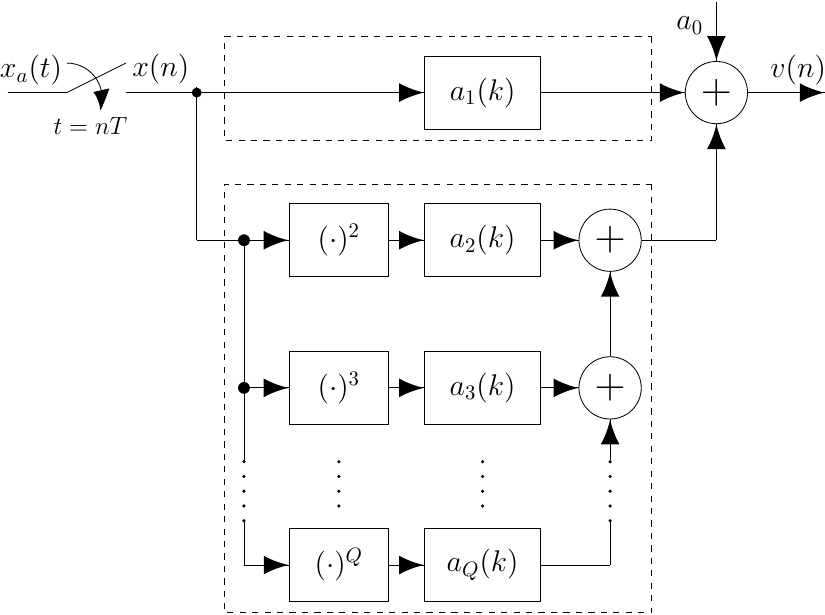}
\par\end{centering}
\caption{Digital pre-sampling Hammerstein model (see Footnote 2) with the upper (lower) dashed
box indicating the linear branch (nonlinear branches).}
\label{Flo:DTS} 
\end{figure}

In the digital pre-sampling Hammerstein model (illustrated in Fig. \ref{Flo:DTS}), $v(n)$ is generated via a memory polynomial \mbox{according} to
\begin{equation}
v(n)=a_{0}+\sum_{k=0}^{D}a_{1}(k)x(n-k)+\sum_{p=2}^{Q}\sum_{k=0}^{D}a_{p}(k)x^{p}(n-k),\label{eq:signal-model}
\end{equation}
where $a_{0}$ is a constant (offset), $a_{1}(k)$ is the impulse response of a linear-distortion discrete-time filter, and $a_{p}(k)$, $p{=}2,3,\ldots,Q$, are the impulse responses of the nonlinear-distortion discrete-time filters. To keep the notation simpler, all filters are here of equal order $D$, but they can generally have different orders. As seen in Fig. \ref{Flo:DTS}, this model corresponds to sampling at the input (pre-sampling) followed by nonlinarities generated in the digital domain. In the special case, when $a_{p}(k)$ are constants (zeroth-order filters), it can equivalently be described as nonlinearities generated in the analog domain followed by sampling since the operations then commute. However, when $a_{p}(k)$ correspond to general filters and the nonlinearities' bandwidths exceed the Nyquist band, one needs to incorporate interpolation for more accurate modeling of the analog distortion (see the extension in Section \ref{sec:analog}). Further, the signal contains quantization noise, but it is here excluded from the mathematical expressions for the sake of simplicity.\footnote{{Signal quantization is included in all evaluations in the examples of Sections III and IV.}}

Before proceeding, it is noted that, to model the signal as in \eqref{eq:signal-model} in a practical system, the parameters $a_0$ and $a_{p}(k)$, $k{=}0,1,\ldots,D$, $p=1,2,\ldots,Q$, need to be estimated. Several methods are available for this purpose \cite{Chen_95,Mao19}. However, the focus of this paper is to assess and compare the performance and complexity of the Hammerstein and proposed linearizers, not to estimate model parameters. To this end, the model in \eqref{eq:signal-model} is used for generating distorted training and evaluation signals. It is stressed though that the proposed linearizers do not assume that the distorted signal is in the form of \eqref{eq:signal-model}. It is also emphasized that the pre-sampling distortion model (and the post-sampling distortion model discussed in Section~IV) is a digital model of aggregate nonlinearities occurring after sampling (before sampling in the post-sampling case). It does not require a physical implementation in the analog domain, but facilitates identification and compensation in the digital domain. 

\color{black} 
\begin{figure}[tb]

\begin{centering}
\includegraphics[scale=0.58]{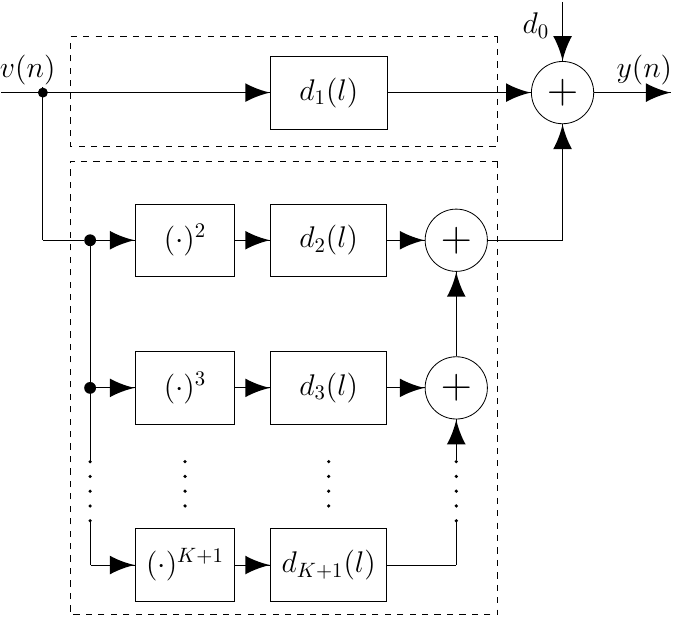}
 
\par\end{centering}
\caption{Hammerstein linearizer with the upper (lower) dashed box indicating the linear branch (nonlinear branches).}
\label{Flo:Hammerstein-scheme} 
\end{figure}

\subsection{Hammerstein Linearizer}\label{sec:Hammerstein-Linearizer}
Given the distorted signal $v(n)$, the linearization amounts to generating a compensated signal, say $y(n)$, in which the distortion has been suppressed (ideally removed). In the conventional Hammerstein linearizer, illustrated in Fig. \ref{Flo:Hammerstein-scheme}, $y(n)$ is generated in the same way as the distorted signal is modeled. That is,
\begin{equation}
y(n)=d_{0}+\sum_{l=0}^{M}d_{1}(l)v(n-l)+\sum_{p=2}^{K+1}\sum_{l=0}^{M}d_{p}(l)v^{p}(n-l),\label{eq:hammerstein_linearizer}
\end{equation}
where $d_{0}$ is a constant (offset), $d_{1}(l)$ is the impulse response of a linear-branch filter, and $d_{p}(l)$, $p{=}2,3,\ldots,K+1$, are the $K$ impulse responses of the $K$ nonlinear-branch filters. Here, $M$ represents the filter order (memory depth). Again, for notation simplicity, all filters have the same order, but they can generally be different. In an implementation, this scheme requires $(M{+}1)(K{+}1){+}K$ multiplications and $(M{+}1)(K{+}1)$ two-input additions per corrected output sample. It involves $(M{+}1)(K{+}1)$ multiplications for generating the filtered versions of $v^{p}(n)$, and $K$ multiplications for generating all $v^{p}(n)$. 

\section{Proposed Linearizer for the Pre-Sampling Distortion Model}
\label{sec:Proposed-Linearizer}

\begin{figure}[t]
	\begin{centering}
		\includegraphics[scale=0.58]{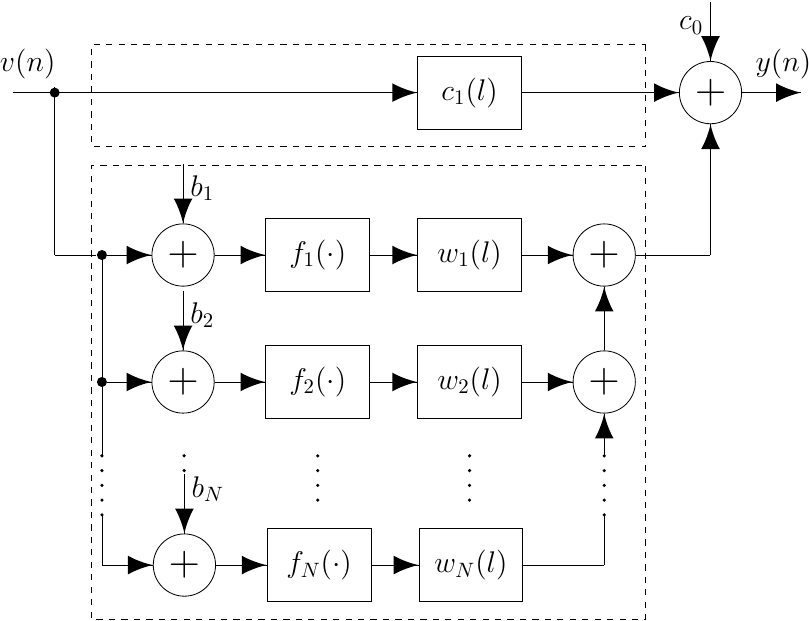} 
		\par\end{centering}
	\caption{Proposed linearizer with the upper (lower) dashed box indicating the
		linear branch (nonlinear branches).}
	\label{Flo:proposed-scheme1} 
\end{figure}

In the proposed linearizer for the pre-sampling model, shown in Fig. \ref{Flo:proposed-scheme1}, the compensated signal $y(n)$ is generated
as
\begin{equation}
y(n)=c_{0}+\sum_{l=0}^{M}c_{1}(l)v(n-l)+\sum_{m=1}^{N}\sum_{l=0}^{M}w_{m}(l)u_{m}(n-l),\label{eq:prop_linearizer}
\end{equation}
where $M$ denotes the filter order {of $c_{1}(l)$ and of the $N$ nonlinear-branch filters $w_{m}(l)$, $m=1,2,\ldots,N$, $c_{0}$ is a constant offset, and the terms $u_{m}(n)$ are}
\begin{equation}
u_{m}(n)=f_{m}(v(n)+b_{m}),
\end{equation}
with $f_{m}$ representing nonlinear operations. Specifically, $f_{m}(v)$ are chosen as either the modulus $|v|$ or the ReLU operation $\max\{0,v\}$ due to their simplicity and reduced complexity in hardware implementation \cite{Tarver_2019,deiro23} (also see Section \ref{sec:alternative-nonlinearities}). Additionally, the bias values $b_{m}$, for $m{=}1,2,\ldots,N$, are selected to be uniformly distributed within the range $[-b_{\max},b_{\max}]$ where the optimal value for $b_{\max}$ is determined as detailed later in Section \ref{subsec:Design}. Hence, the bias values $b_{m}$ are chosen as
\begin{equation}
b_{m}=-b_{\max}+\frac{2(m-1)b_{\max}}{N-1}, \quad m=1,2,\ldots,N.\label{eq:bm}
\end{equation}

\begin{figure}[t]
\begin{centering}
\includegraphics[scale=0.58]{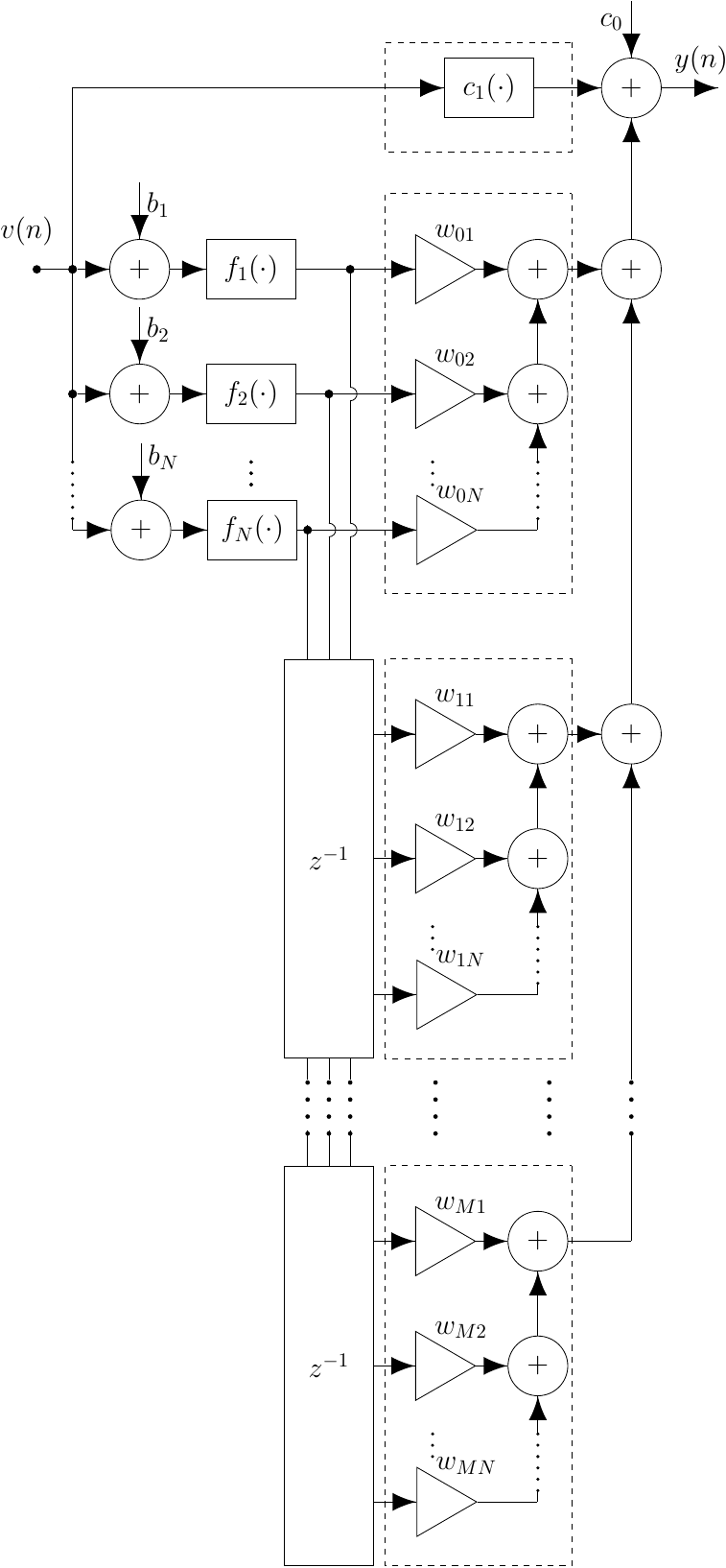} 
\par\end{centering}
\caption{Equivalent implementation of the proposed linearizer in Fig. \ref{Flo:proposed-scheme1}, utilized in the proposed design.}
\label{Flo:proposed-scheme2} 
\end{figure}

\subsection{Implementation Complexity}\label{sec:Proposed-Linearizer-Complexity}
In the implementation, the proposed scheme in \eqref{eq:prop_linearizer} requires only $(M{+}1)(N{+}1)$ multiplications per sample, and $(M{+}1)(N{+}1){+}N$ two-input additions, including the $N$ bias additions. Notably, the proposed linearizer requires $K$ multiplications less than the Hammerstein model at the expense of $K$ extra additions, when they have the same number of branches ($N{=}K$). Hence, in particular for cases where a small $M$ is sufficient, the multiplication complexity is substantially lower for the proposed linearizer (also see Section~\ref{subsec:complexity1}). Further, since multiplications generally require substantially more power than additions in an implementation~\cite{parhi2007vlsi, koren2001computer}, the proposed linearizer will have a lower implementation complexity.

Moreover, a significant additional advantage of the proposed linearizer is that it eliminates the need for data quantization before filtering. Conversely, in the Hammerstein linearizer, data quantizations must be performed at the outputs of the nonlinearities (i.e., before the filtering) to prevent excessively long and costly internal word lengths. These quantizations introduce quantization errors (quantization noise), which are scaled by the energy of the corresponding filters' impulse responses $d_{k}(l)$ which equals the sum of the squares of the impulse response values \cite{Jackson_96, Wanhammar_13}. Consequently, the impulse response values must be kept small to avoid significant noise amplification, which would degrade the output SNDR. Alternatively, longer internal word lengths need to be used, which also increases the implementation cost. The proposed scheme overcomes this problem since all quantizations are carried out after the filtering operations, thereby avoiding noise amplification from the quantizations to the output.

\subsection{Proposed Design}
\label{subsec:Design}

For design purposes, we make use of the equivalent structure in Fig. \ref{Flo:proposed-scheme2} where $w_{lm}{=}w_{m}(l)$. The design of the proposed linearizer then amounts to determining the parameters $c_{0}$, $c_{1}$, $w_{lm}$, and $b_{m}$ so that the output signal $y(n)$ closely approximates the desired signal $x(n)$ in some sense,
which is here assumed to be the least-squares sense. To this end, the design procedure described below is proposed, which extends our frequency-independent linearizer in \cite{deiro23} to accommodate frequency dependency. It is assumed that the signal is normalized so that its modulus is bounded by one.
\begin{enumerate}
\item Generate a set of $R$ reference signals $x_{r}(n)$ and the corresponding distorted signals $v_{r}(n)$, $r{=}1,2,\ldots,R$, using a signal model as in \eqref{eq:signal-model} or measured data. 
\item Generate a set of $S$ uniformly distributed values of $b_{\max}\in[b_{l},b_{u}]$. 
We have experimentally observed that $b_{l}{=}0.5$ and $b_{u}{=}1.5$ are appropriate values when the signal is within the range $[-1,1]$.
\item For each specified value of $b_{\max}$, the corresponding $b_{m}$ values are calculated as in \eqref{eq:bm}. Then, minimize the cost function $E$ given by
\begin{equation}
E=\sum_{r=1}^{R}\sum_{n=1}^{L}\left(y_{r}(n)-x_{r}\Bigl(n-\frac{M}{2}\Bigr)\right)^{2},\label{eq:E}
\end{equation}
where $L$ denotes the data length and $M/2$ compensates for the delay of the linearization filters\footnote{It is assumed in the expression $n-M/2$ that $M$ is even, but odd $M$ can also be handled by replacing $M/2$ with $(M-1)/2$ or $(M+1)/2$.}. The coefficients $c_{0}$, $c_{1}$, and $w_{l,m}$ (for $l{=}0,1,\ldots,M$, and $m{=}1,2,\ldots,N$) in \eqref{eq:prop_linearizer} are computed using matrix inversion, incorporating $L_{2}$-regularization to avoid large parameter values and to prevent ill-conditioned matrices. Specifically, let $\mathbf{w}$ be a $([(M{+}1)(N{+}1){+}1]\times1)$ column vector containing all coefficients $w_{l,m}$, along with $c_{1}(l)$ and $c_{0}$. Let $\mathbf{A}_{r}$ be an $(L\times[(M{+}1)(N{+}1){+}1])$ matrix, where each column contains the $L$ samples of $u_{r,m}(n{-}l)$ for $l{=}0,1,\ldots,M$ and $m{=}1,2,\ldots,N$, the $L$ input samples $v_{r}(n{-}l)$ for $l{=}0,1,\ldots,M$, and $L$ ones (corresponding to the constant $c_{0}$). Minimizing $E$ in
\eqref{eq:E}, in the least-squares sense, yields the solution (see Appendix \ref{sec:LS_Solution})
\begin{equation}
{\bf w}={\bf A}^{-1}{\bf b},
\label{eq:w7}
\end{equation}
where
\begin{equation}
{\bf A}=\lambda{\bf I}_{(M+1)(N+1)+1}+\sum_{r=1}^{R}{\bf A}_{r}^{\top}{\bf A}_{r},\quad{\bf b}=\sum_{r=1}^{R}{\bf A}_{r}^{\top}{\bf b}_{r},
\label{eq:w8}
\end{equation}
with ${\bf A}_{r}^{\top}$ being the transpose of ${\bf A}_{r}$, and ${\bf b}_{r}$ being an $L\times1$ column vector containing the $L$ samples $x_{r}(n{-}M/2)-v_{r}(n{-}M/2)$. It is noted here that $x_{r}(n{-}M/2)-v_{r}(n{-}M/2)$ is used in ${\bf b}_{r}$ instead of $x_{r}(n{-}M/2)$, in order to compute small values of the linear-branch filter coefficients. That is, we replace $c_{1}(l)v(n{-}l)$ in \eqref{eq:prop_linearizer} with $v(n{-}M/2)+\Delta c_{1}(l)v(n{-}l)$ and then compute the value of $\Delta c_{1}$. In this way, all parameters to be computed in the least-squares design are small (zero in the ideal case with no distortion).\footnote{{The paper focuses on weakly nonlinear systems where the nonlinearities are much smaller than the desired signal and the model parameters are small, which is typically the case in analog circuits with undesired nonlinearities. In this case, the linearizers can also have relatively small coefficients, which in the proposed design are obtained through matrix inversion with $L_2$-regularization. In systems with larger model parameters, the linearizers may need larger coefficients which can also be obtained through the proposed design by using a smaller $\lambda$ for the $L_2$-regularization.}} Further, $\lambda{\bf I}_{(M+1)(N+1)+1}$ is a diagonal matrix with small diagonal entries $\lambda$ for the $L_{2}$-regularization. The linearized output (treated as a row vector) ${\bf y}_{r}{=}y_{r}(n)$, $n{=}1,2,\ldots,L$, is given by
\begin{equation}
{\bf y}_{r}{=}{\bf v}_{r}+{\bf w}^{\top}{\bf A}_{r}^{\top},
\end{equation}
where ${\bf v}_{r}{=}\sum_{l{=}0}^{M}v_{r}(n{-}l)$, $n{{{=}}}1,2,\ldots,L$ is also a row vector. 
\item Select the best of the $S$ solutions above. 
\item Evaluate the linearizer over a large set of signals, say $R^{(eval)}$, where $R^{(eval)}\gg R$. 
\end{enumerate}

{\bf Remark 1: }The proposed design uses $R$ reference signals. Once these reference signals have been collected, the design amounts to a matrix inversion for a given set of bias values. Hence, the design time (and thus convergence time) is mainly determined by the time it takes to collect the $R$ reference signals. This is however common for all linearizers, which require the same amount of data to achieve the same linearizer performance, and is thus not specific for our proposal. The additional aspect of finding the bias values can in practice be carried out only once, and one can then use those bias values when the model and model parameters change. This is because we have observed that the linearizer performance {has a low sensitivity} to exact bias values, as long as they are linearly spaced between the approximate minimum and maximum of the signal.\footnote{{To assess the bias sensitivity, we changed the optimal $b_{\max}$ in the examples by $\pm 3$ and $\pm 5$ percent. This caused a mean SNDR reduction of less than $1$ and $2$ dB, respectively.}} Taking this into account, the proposal has the same design time as the Hammerstein linearizer, but offers a lower linearizer implementation complexity. It is also stressed again that the proposal avoids the costly and time-consuming iterative nonconvex optimization that is traditionally used when training neural networks.

{\bf Remark 2: }In the examples later on where synthetic data is used (Examples 1--5), the reference signals are known (e.g. pilot sequences). It is appropriate to use such signals when investigating the linearization capability of the linearizers, both for the proposed one and Hammerstein. In practice, the reference signals may not be completely known but have to be estimated. As long as the signal estimates are sufficiently accurate, the proposed design still works well. This will be illustrated in Example 6 where sinusoids with unknown amplitudes and phases are used. In general, existing methods for estimation of multi-sine signals can be used in this context.

{{\bf Remark 3: }In the proposed method, the bias values are optimized over a discrete set of values, which means that the overall solution is not guaranteed to be globally optimal, even if each solution with fixed bias values so is. However, also for a regular optimization, where the bias values and coefficients are jointly optimized, one cannot guarantee global optimality as the problem is then not convex. Hence, only local optimality can be guaranteed. Starting with the proposed optimized solutions in the example sections, we have also carried out further joint optimizations which do not improve the results. This shows that the proposed design yields at least locally optimal solutions, which are also good solutions as they outperform the Hammerstein linearizers, whose optimized solutions are guaranteed to be globally optimal since they lack bias values.}


\subsection{Simulations and Results}
\label{subsec:Simulations-and-results}

\begin{figure}[tb]
\begin{centering}
\includegraphics[scale=1.00]{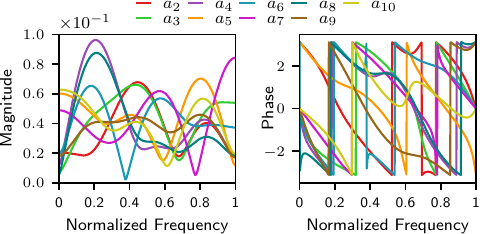} 
\par\end{centering}
\caption{Magnitude and phase response of $a_{p}(k)$, for $p=2,\cdots,10$, in \textit{Example~1}.}
\label{Flo:mag_phase} 
\end{figure}

For the evaluations and comparisons, we assume a distorted signal $v(n)$ with a distortion filter order of $D{=}6$. This signal is modeled as described in \eqref{eq:signal-model} with parameters\footnote{We have also considered cases where $a_{0}$ and $a_{1}(k)$ were randomly generated (additional small offset and linear distortion), but the results were practically the same.}: $a_{0}{=}0$, $a_{1}(k){=}[0,0,0,1,0,0,0]$, and $a_{p}(k)$ for $k{=}0,1,\ldots,D$ and $p{=}2,3,\ldots,Q$, with $Q{=}10$, randomly generated with frequency responses as depicted in Fig. \ref{Flo:mag_phase}. We imposed a constraint to have a mean SNDR around $30$ dB for the set of distorted signals. The SNDR in dB for a real reference signal $x(n)$ and its distorted signal $v(n)$, computed over $L$ samples, is
\begin{equation}
	\text{SNDR}=10\log_{10}\Biggl(\frac{\sum_{n=0}^{L-1} x^2(n)}{\sum_{n=0}^{L-1} (x(n)-v(n))^2} \Biggr)\text{ [dB]}.
\end{equation}

\textit{Example 1: }We consider the multi-tone signal  \label{ex:1}
\begin{equation}
x(n)=G\times\sum_{k=1}^{31}A_{k}\sin(\omega_{k}n+\alpha_{k}),\label{eq:OFDM}
\end{equation}
where $A_{k}{=}1$ for all $k$, and $\alpha_{k}$ are randomly chosen from $\{\pi/4,-\pi/4,3\pi/4,-3\pi/4\}$, which corresponds to quadrature phase shift keying (QPSK) modulation. The frequencies $\omega_{k}$ are given by
\begin{equation}
\omega_{k}=\frac{2\pi k}{64}+\Delta\omega,\label{eq:omegak}
\end{equation}
in which case the signal corresponds to the quadrature (imaginary) part of $31$ active subcarriers in a $64$-subcarrier OFDM signal with a random frequency offset $\Delta\omega$. In the design and evaluation we use, respectively, $R{=}50$ and $R^{(eval)}{=}5000$ signals with randomly generated frequency offsets assuming uniform distribution between $-\pi/64$ and $\pi/64$, quantized to $12$ bits, and of length $L{=}8192$. The gain $G$ is selected so that the distorted signals are below one in magnitude. For the $L_{2}$-regularization, a linear search is used to find the value $\lambda\in[10^{-10},10^{-1}]$ that best fits each instance, meaning each combination of $b_{\text{max}}$, $M$, and $N$. The Hammerstein linearizer has been designed in the same way but without the bias values. The constraint was to select the optimal $\lambda$-value\footnote{\label{foo:lambda}The optimal $\lambda$-value is defined here as the value that results in the smallest error in \eqref{eq:E} while avoiding the matrix $\lambda{\bf I}+\mathbf{A}^{\top}\mathbf{A}$ to be \mbox{ill-conditioned} and ensuring that all the entries in $\mathbf{w}$ are within the range $[-1,1]$.}, which was primarily used to regulate the Hammerstein linearizer to avoid large parameter values and large noise amplification. The proposed linearizer does not have noise amplification and thus allows larger parameter values, but the magnitudes of those values are nevertheless below unity in magnitude, even with a relaxed $L_{2}$-regularization.

Further, we have considered both the modulus and ReLU as nonlinear operations, as well as combinations (modulus in some branches, ReLU in the other branches), but the different options resulted in practically the same performance. The results presented in the examples are for the modulus operation.

\begin{figure}[tb]
\begin{centering}
\includegraphics[scale=1.15]{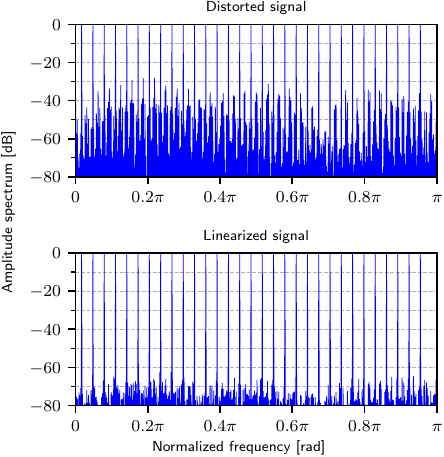}  
\par\end{centering}
\caption{Spectrum before and after linearization for a multi-sine signal using the proposed linearizer with a filter order of $M{=}6$ and $N{=}12$ nonlinear branches (\textit{Example 1}).}
\label{Flo:Spectrum1} 
\end{figure}

\begin{figure}[tb]
	\begin{centering}
		\includegraphics{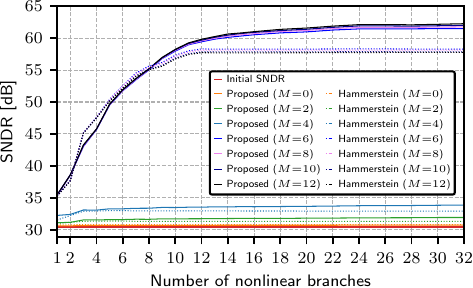} 
		\par\end{centering}
	\caption{SNDR versus number of nonlinear branches in \textit{Example 1}.}
	\label{Flo:SNDRvsBranch} 
\end{figure}

Figure \ref{Flo:Spectrum1} shows the spectrum before and after linearization for one of the signals using the proposed linearizer with a filter order of $M{=}6$ and $N{=}12$ nonlinear branches. Figure \ref{Flo:SNDRvsBranch} plots the mean SNDR over 5000 signals for each linearizer instance (with an SNDR variance of approximately 0.5 dB for all instances) versus the number of branches for the proposed and Hammerstein linearizers.\footnote{{For all designs, the optimized parameter values and arithmetic-operation results are quantized to 14 bits in the evaluations. This provides a practical trade-off between hardware cost and numerical accuracy. Increasing precision beyond this yields negligible SNDR improvements.}}

The signal-to-noise ratio (SNR) is approximately 65 dB without distortion, and the SNDR is around 30 dB for the distorted signals before linearization. As seen in Fig. \ref{Flo:SNDRvsBranch}, the SNDR approaches 62 dB for the proposed linearizer, thus an improvement by 32 dB which corresponds to more than five bits improvement when using a linearizer filter order of or above six, which is required here because the distortion filter order is six. For the Hammerstein linearizer, the SNDR approaches only about 58\,dB. The difference between the SNDR values (62\,dB and 58\,dB) and the bound of 65\,dB (since $\textnormal{SNDR} \leq \textnormal{SNR}$) can be attributed to the $L_{2}$-regularization. The SNDR gap of 4\,dB is due to the fact that the $L_{2}$-regularization has less impact on the proposed linearizer, which is because it has relatively small coefficients even without $L_{2}$-regularization.\footnote{{The performance gap of 4 dB can be decreased by using a smaller regularization parameter $\lambda$ but it also comes with larger coefficient values and increased implementation cost as more bits internally are then required in the Hammerstein linearizer implementation. As discussed in \mbox{Section~\ref{sec:Proposed-Linearizer-Complexity}}, this is due to noise amplification which is present in the Hammerstein linearizer but not in the proposed linearizer. To exemplify, in \textit{\textit{\hyperref[ex:1]{Example~1}}} with 24 branches and $M{=}6$, reducing $\lambda$ by a factor of 250 for the Hammerstein linearizer, causes the maximum coefficient magnitude to increase by a factor of 56 compared to the optimal $\lambda$ (see Footnote~\ref{foo:lambda}). Further, to reduce the performance gap to, e.g., about 1~dB, 22~bits are required instead 14~bits, which is sufficient in the proposed linearizer.}} This implies that the Hammerstein linearizer has a lower peak performance (about 4\,dB in this example) than the proposed linearizer for the same $L_{2}$-regularization.

\begin{figure}[tb]
\begin{centering}
\includegraphics[scale=1.15]{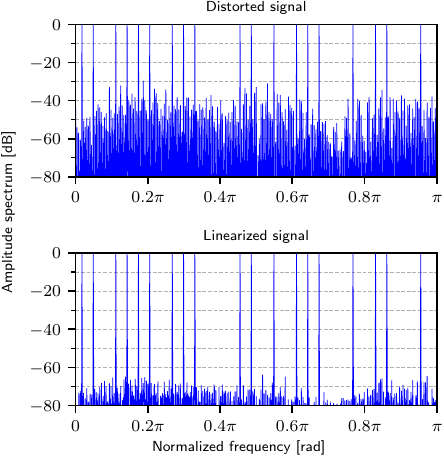} 
\par\end{centering}
\caption{Spectrum before and after linearization for a multi-sine signal with null subcarriers using the proposed linearizer with a filter order of $M{=}6$ and $N{=}12$ nonlinear branches (\textit{Example 2}).}
\label{Flo:Spectrum2} 
\end{figure}

\begin{figure}[tb]
\begin{centering}
\includegraphics[scale=1.0]{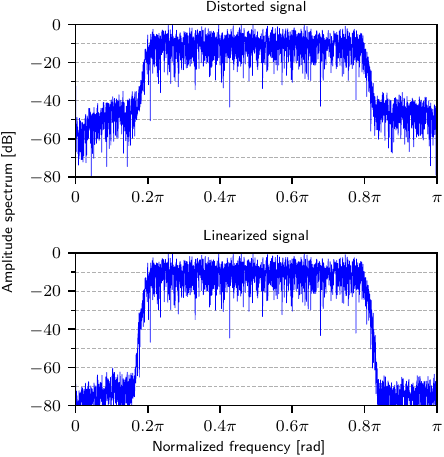}  
\par\end{centering}
\caption{Spectrum before and after linearization for a bandpass filtered white-noise signal using the proposed linearizer with a filter order of $M{=}6$ and $N{=}12$ nonlinear branches (\textit{Example 2}).}
\label{Flo:Spectrum3} 
\end{figure}

\textit{Example 2: }To further illustrate the robustness of the proposed linearizer designed in \textit{Example 1}, we have also evaluated it for the same type of multi-sine signal as in \textit{Example 1} but with some subcarriers set to zero, and for a bandpass filtered white-noise signal covering 60\% of the Nyquist band. As illustrated in Figs. \ref{Flo:Spectrum2} and \ref{Flo:Spectrum3} for one of each of these signals, essentially the same result is obtained.\footnote{The observed out-of-band spectral reduction in Fig. \ref{Flo:Spectrum3} is due to linearization of a distorted bandpass filtered white-noise signal. In this case the linearizer suppresses the out-of-band distortion down to the noise floor which emanates from data quantization.} Less than 1 dB SNDR degradation compared to the linearized signals considered in \textit{Example 1} is observed.

{{\bf Remark 4: }It is noted that the multi-sine signals in \textit{Examples~1~and~2} correspond to the imaginary part of an OFDM signal (as real signals are considered in the paper) with QPSK (i.e., 4-QAM) modulation,  whereas the bandpass filtered white-noise signals in \textit{Example~2} resemble higher-order QAM-modulated signals. Additional simulations with 16-QAM and 64-QAM show practically the same performance, confirming that the method generalizes well across general wideband communication signals.}	

\begin{figure}[t]
	\begin{centering}
		\includegraphics[scale=1.00]{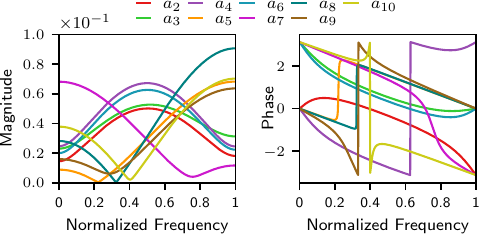} 
		\par\end{centering}
	\caption{Magnitude and phase responses of $a_{p}(k)$, for $p=2,\cdots,10$,
		in \textit{Example 3}.}
	\label{Flo:mag_phase2} 
\end{figure}

\begin{figure}[t]
	\begin{centering}
		\includegraphics{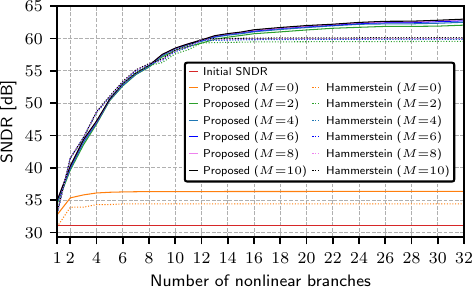} 
		\par\end{centering}
	\caption{SNDR versus number of nonlinear branches in \textit{Example 3}.}
	\label{Flo:SNDRvsBranch2} 
\end{figure}

\textit{Example 3: } This example presents results similar to those in Fig. \ref{Flo:SNDRvsBranch} for \textit{Example 1}, but with 5000 OFDM signals with QPSK modulation exhibiting second-order filtered distortion terms ($D{=}2$) with the parameters $a_{0}{=}0$, $a_{1}{=}[0,1,0]$, and randomly generated $a_{p}$ for $p{=}2,3,\ldots,Q$, with $Q{=}10$, and whose frequency responses are as shown in Fig. \ref{Flo:mag_phase2}. In this case, the signal can be well linearized with an order of $M\geq2$, due to the simpler second-order filtered distortion terms. This is seen in Fig. \ref{Flo:SNDRvsBranch2}.


\subsubsection{Implementation Complexity}
\label{subsec:complexity1}

\begin{figure}[t]
	\begin{centering}
		\includegraphics{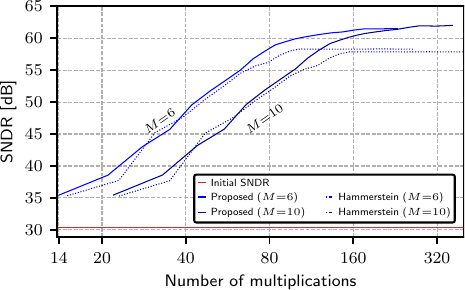} 
		\par\end{centering}
	\caption{SNDR versus number of multiplications in\textit{ Example 1}.}
	\label{Flo:SNDR_versus_multiplications0} 
\end{figure}

\begin{figure}[t!]
	\begin{centering}
		\includegraphics[scale=0.98]{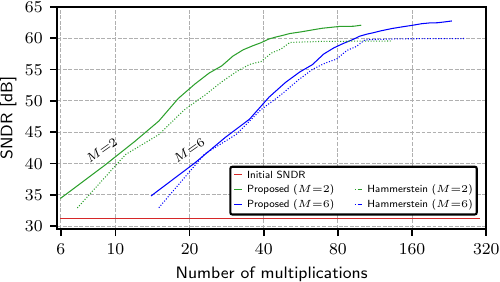} 
		\par\end{centering}
	\caption{SNDR versus number of multiplications in\textit{ Example 3}.}
	\label{Flo:SNDR_versus_multiplications} 
\end{figure}

Figures \ref{Flo:SNDRvsBranch} and \ref{Flo:SNDRvsBranch2} show the SNDR improvement versus the number of branches in the linearizer. However, as discussed in Sections~\ref{sec:Signal-model} and \ref{sec:Proposed-Linearizer}, with $N$ nonlinear branches, the Hammerstein linearizer (with $N{=}K$) requires $K$ additional multiplications compared to the proposed scheme, which has $K$ additional additions, but since multiplications are generally more expensive to implement than additions~\cite{parhi2007vlsi, koren2001computer}, the implementation cost will be lower for the proposal. The proposed linearizer is thus more efficient for the same number of nonlinear branches, especially when $M$ is small and the relative cost of multiplications is relatively high. This efficiency is demonstrated in Figs. \ref{Flo:SNDR_versus_multiplications0} and \ref{Flo:SNDR_versus_multiplications}, which plot the SNDR versus the number of multiplications for both methods in \textit{Example 1} and \textit{Example 3}, respectively. It is observed that the proposed linearizer clearly outperforms the Hammerstein linearizer for a small $M$ ($M{=}2$ in Fig. \ref{Flo:SNDR_versus_multiplications}) whereas the two methods have comparable performance for a larger $M$ ($M{=}6$ in Figs. \ref{Flo:SNDR_versus_multiplications0} and \ref{Flo:SNDR_versus_multiplications} and $M{=}10$ in Fig. \ref{Flo:SNDR_versus_multiplications0}). {For instance, in \textit{Example~3} with $M{=}2$, the Hammerstein linearizer requires 43 multiplications to reach $57~\text{dB}$ (Fig.~\ref{Flo:SNDR_versus_multiplications}), whereas the proposed linearizer achieves the same performance with only 30 multiplications, corresponding to a saving of $13/43 {\approx} 30\%$.} It is also observed again that the proposed linearizer achieves up to 4\,dB higher SNDR than the Hammerstein linearizer, as seen in Figs.~\ref{Flo:SNDRvsBranch} and \ref{Flo:SNDRvsBranch2}--\ref{Flo:SNDR_versus_multiplications}, due to the $L_{2}$-regularization. {Finally, although the results here focus on distortion orders $D{=}6$ and $D{=}2$, it should be noted that the largest savings (about $50\%$) of the proposed linearizer over the Hammerstein linearizer occur in the memory-independent case ($D{=}0$), which allows a memoryless linearizer ($M{=}0$), as previously shown in~\cite{deiro23}.}

 As mentioned in the introduction, it is also noted that both the proposed and Hammerstein linearizers are substantially more efficient than existing neural-network-based linearizers which require several hundreds or even thousands of multiplications to correct each output sample, even for simpler and more narrowband signals and for more modest SNDR improvements \cite{Xu2019DeepLearningADC,Chen20,DENG_202063,Peng_2021,Chen21,Fayazi21,Zhai22,Zhifei24,Peng_2024}. For example, the proposed linearizer in \textit{Example 3} (Figs. \ref{Flo:SNDRvsBranch2} and \ref{Flo:SNDR_versus_multiplications}), when it starts reaching its SNDR saturation level, requires about 40 multiplications for $M{=}2$. (Increasing $M$ and $N$ further only offers a modest SNDR improvement at the cost of a higher complexity). Correspondingly, in \textit{Example 1} (Figs.  \ref{Flo:SNDRvsBranch} and \ref{Flo:SNDR_versus_multiplications0}), about 80 multiplications are required for $M{=}6$ (recall that $M{=}6$ and $M{=}2$ correspond to the distortion filter order, $D{=}6$ and $D{=}2$, in \textit{Example 1} and \textit{Example 3}, respectively). This is about an order-of-magnitude lower complexity than for the existing neural-network-based linearizers.

\section{Proposed Linearizer for the Post-Sampling Distortion Model}
\label{sec:analog}

\begin{figure}[!t]

	\begin{centering}
	\includegraphics[scale=0.58]{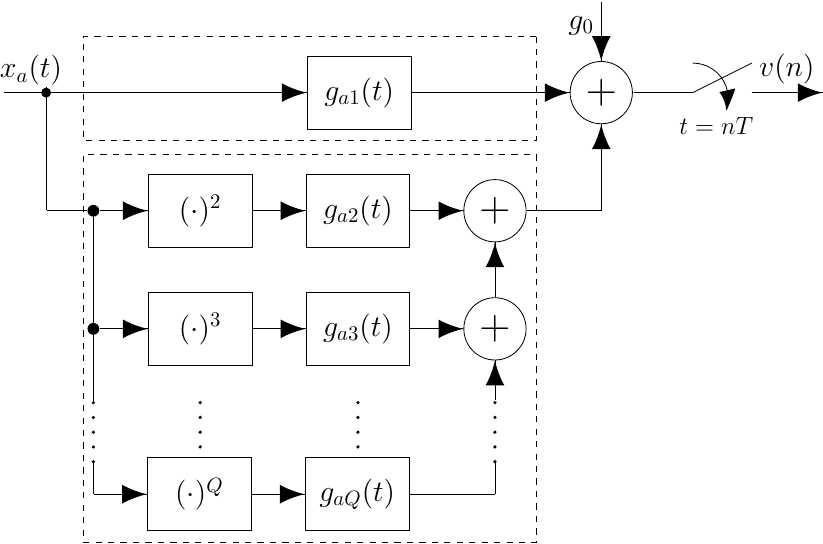}
	
	\par\end{centering}
	\caption{Analog post-sampling Hammerstein model with the upper (lower) dashed box indicating the linear branch (nonlinear branches).}
 \label{Flo:CT} 
\end{figure}

\begin{figure}[t]
	\begin{centering}
		\includegraphics[scale=0.48]{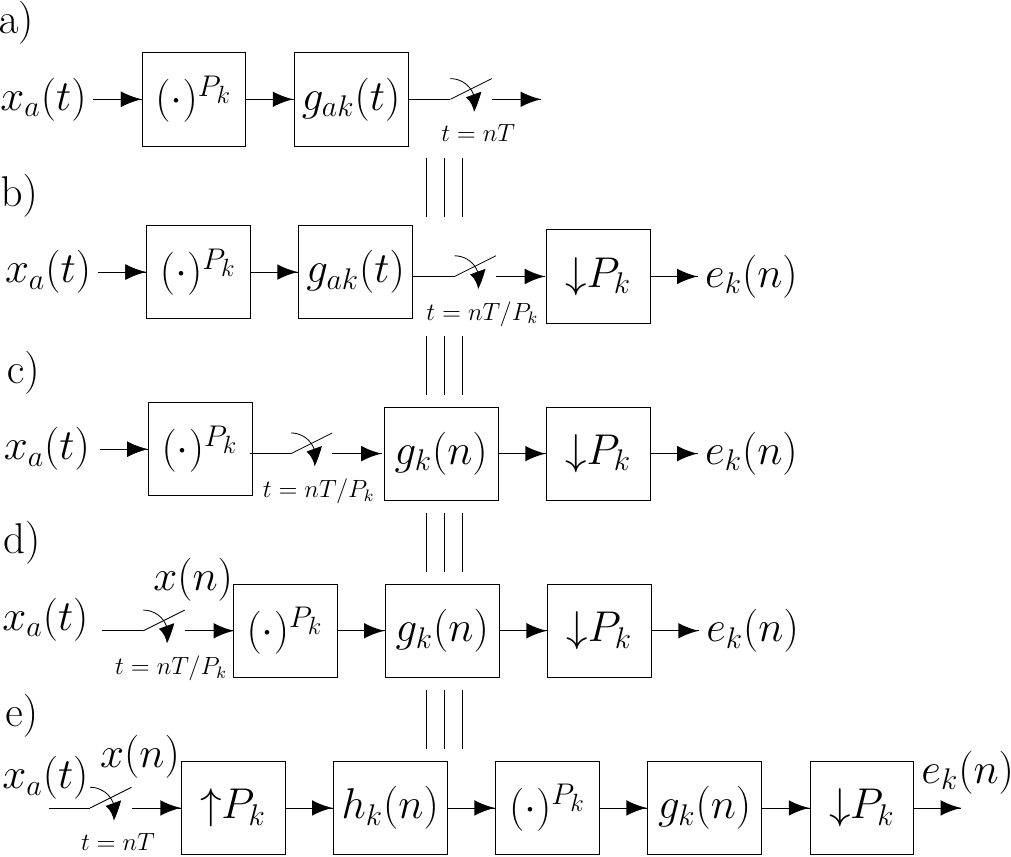}
		\par\end{centering}
	\caption{Derivation of the discrete-time equivalence of branch $k$ in Fig. \ref{Flo:CT}.}
	\label{Flo:signal-model}
\end{figure}

To linearize post-sampling analog distortion, the proposed linearizer in Section \ref{sec:Proposed-Linearizer} is extended by incorporating interpolation. The point of departure is then the analog Hammerstein model in Fig. \ref{Flo:CT} where the distorted digital signal $v(n)$ is obtained by sampling the corresponding distorted analog signal. Here, even if $x_{a}(t)$ is bandlimited to the Nyquist band, the nonlinear distortion is not, because the nonlinearities $(\cdot)^{p}$ widen the spectrum by a factor of $p$. However, as discussed in the introduction, it is still possible to recover the desired sequence $x(n)$ from $v(n)$ \cite{Tsimbinos_98}, which in practice can be carried out by utilizing interpolation in the linearization \cite{Vansebrouck_2017}. To this end, we will make use of a discrete-time equivalence to the structure in Fig. \ref{Flo:CT}, which was not considered in \cite{Vansebrouck_2017}. The discrete-time equivalence is derived through the steps shown in Fig. \ref{Flo:signal-model}(a)-(e) for one branch, which is further explained below.

\subsection{Equivalent Discrete-Time Model}\label{subsec:Equivalent-Discrete-Time-Model}
\begin{figure}[tb]
	\begin{centering}
		\includegraphics[scale=0.60]{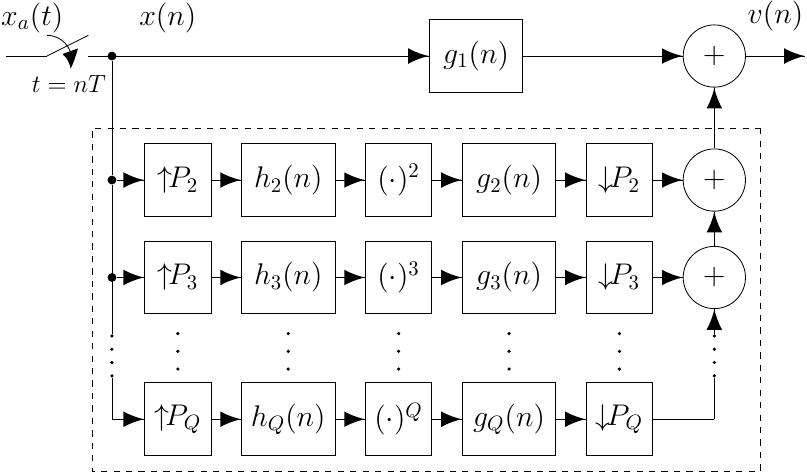} 
		\par\end{centering}
	\caption{Discrete-time equivalence of the scheme in Fig. \ref{Flo:CT}.}
	\label{Flo:discrete-time-equivalence2} 
\end{figure}

Starting with the branch scheme in Fig. \ref{Flo:signal-model}(a), we first replace the sampling at the output with a $P_{k}$-fold faster sampler\footnote{It is assumed here that $P_k{=}k$, for $k{=}2,3,\ldots,Q$, (Fig. \ref{Flo:CT}), but in general, $P_k$ can take on values from a set of integer values.} followed by a downsampler that discards the redundant samples, resulting in Fig. \ref{Flo:signal-model}(b). Next, we note that the sampling at the output of the filter $g_{ak}(t)$ fulfills the sampling theorem for the sampling rate $P_{k}/T$. Hence, as seen in Fig. \ref{Flo:signal-model}(c), the filter $g_{ak}(t)$ followed by the sampler can be replaced by a sampler followed by a digital filter $g_{k}(n)$ having the same frequency response as $g_{ak}(t)$, thus $G_k(e^{j\omega T/P_{k}}){=}G_{ak}(j\omega)$, in the frequency region $\omega\in[0,\pi P_{k}/T]$. As the nonlinearity is static (memoryless), we can then interchange the order of the sampler and nonlinearity, as seen in Fig. \ref{Flo:signal-model}(d). Now, we note that $x_{a}(t)$ is oversampled $P_{k}$ times. Therefore, the input to the nonlinearity can equivalently be obtained through sampling at the original lower rate $1/T$ followed by interpolation by $P_{k}$, the latter being represented by upsampling by $P_{k}$ followed by the discrete-time interpolation filter $h_{k}(n)$. This yields Fig. \ref{Flo:signal-model}(e). Making use of this equivalence, the discrete-time equivalence to the whole scheme in Fig. \ref{Flo:CT} is obtained according to Fig. \ref{Flo:discrete-time-equivalence2}. Here, we have also utilized that, in the linear branch, $g_{a1}(t)$ can be directly modeled by its discrete-time counterpart $g_{1}(n)$ since filtering followed by sampling is equivalent to sampling followed by filtering, provided that only linear operations are involved and the signal is Nyquist sampled. Based on the discrete-time equivalence derived above, the linearizers follow in the same way as in Sections \ref{sec:Signal-model} and \ref{sec:Proposed-Linearizer} as detailed below. The design of the linearizers follows the same procedure as proposed in Section~\ref{sec:Proposed-Linearizer}.

\subsection{Hammerstein Linearizer}\label{subsec:Hammerstein Linearizer-Interp}
\begin{figure}[t]
	\centering{}\raggedright \includegraphics[scale=0.6]{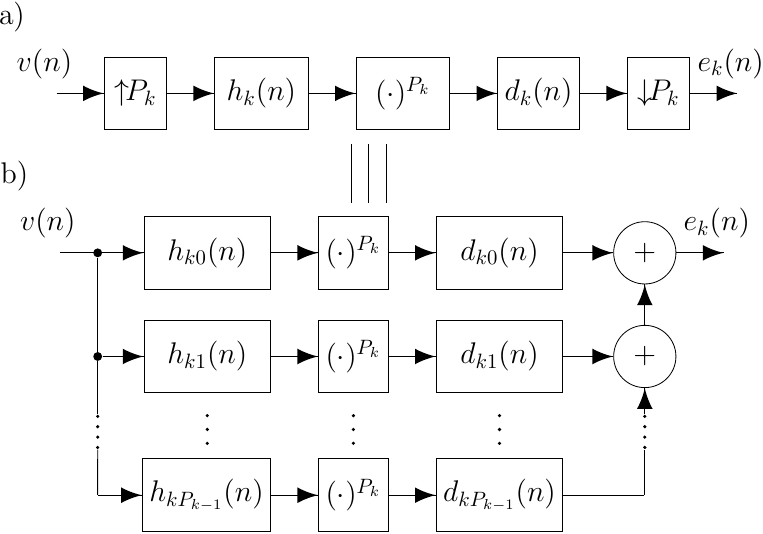}
	\caption{Nonlinear-branch implementation of the Hammerstein linearizer.}
	\label{Flo:hammerstein-scheme_analog} 
\end{figure}

For the Hammerstein linearizer, the compensated signal $y(n)$ is again generated in the same way as the distorted signal is modeled. To avoid filtering operations at the higher sampling rate [see Fig. \ref{Flo:signal-model}(e)] we make use of polyphase decomposition \cite{Vaidyanathan_93}, and write the transfer functions $H_{k}(z)$ and $G_{k}(z)$ according to
\begin{equation}
	H_{k}(z)=\sum_{i=0}^{P_{k}-1}z^{-i}H_{ki}(z^{P_{k}})\label{eq:5}
\end{equation}
and
\begin{equation}
	G_{k}(z)=\sum_{i=0}^{P_{k}-1}z^{i}G_{ki}(z^{P_{k}}),\label{eq:6}
\end{equation}
respectively. Through the use of the corresponding polyphase realizations, each nonlinear branch can then be implemented according to Fig. \ref{Flo:hammerstein-scheme_analog}(b) where all operations are carried out at the input-output sampling rate. Utilizing this, it is observed that the final discrete-time model and linearizer belong to the class of parallel LNL systems, where each branch comprises a filter (here polyphase component), a static nonlinearity $(\cdot)^{P_k}$, and a filter whose coefficients are determined by the ADI's distortion. The main difference from the Hammerstein linearizer in Section \ref{sec:Signal-model} (Fig. \ref{Flo:Hammerstein-scheme}) is that, here, each static nonlinearity $(\cdot)^{P_k}$ is present in $P_k$ polyphase branches and each branch incorporates the interpolation filters' polyphase components [Fig. \ref{Flo:hammerstein-scheme_analog}(b)].

\subsubsection{Implementation Complexity}\label{subsec:Hammerstein Linearizer-Interp-Complexity}
The interpolation filters are excluded in the complexity comparisons as they are common for the Hammerstein and the proposed schemes. Compared with the pre-sampling scheme, the number of multiplications required for the post-sampling Hammerstein linearizer increases because several copies of each static nonlinearity \( (\cdot)^{P_k} \) need to be implemented and the generation of the different nonlinearities can not be shared between the branches as they have different inputs (different polyphase components' outputs). Thus, the number of multiplications required to correct each output sample increases to \( (M+1)\times(K+1)+S(P_{K+1}) \), where
\begin{equation}
    S(P_{K+1}) = \sum_{k=2}^{P_{K+1}} \underset{S_k}{\underbrace{\min(k, M+1) \phi\left\{(\cdot)^k\right\}}}
\end{equation}
represents the number of multiplications required to create all the static nonlinearities $(\cdot)^k$ for the nonlinear branches $k{=}2, \cdots, P_{K+1}$. Here, $\phi\left\{(\cdot)^k\right\}$ represents the minimum number of multiplications required to generate one of the nonlinearities $(\cdot)^k$ in a particular branch $k$, which can be found through an optimal addition-chain exponentiation algorithm \cite{hatem2021}, whereas $S_k$ represents the total nonlinearity-multiplication complexity of branch $k$. Further, when the filter length $M+1$ is smaller than $P_k=k$, $k-M-1$ polyphase components become zero, which explains the term $\min(k, M+1)$ in the summation. To exemplify, the minimum number of multiplications to generate all the nonlinearities \((\cdot)^k\), for $P_{K+1}=10$ and $M{=}2$ ($M+1\geq 10$), is $S(10){=}77\, (177)$, distributed as $S_2{=}2\, (2)$, $S_3{=}6\, (6)$, $S_4{=}6\, (8)$, and so on until $S_{10}{=}12 \, (40)$ with the sequence\footnote{Note that the sequences $(\cdot) \xrightarrow{\!\!\times (\cdot)\!} (\cdot)^2 \xrightarrow{\!\!\times (\cdot)^2\!\!\!} (\cdot)^4 \xrightarrow{\!\!\times (\cdot)^2\!\!\!\!} (\cdot)^6 \xrightarrow{\!\!\times (\cdot)^4\!\!\!\!} (\cdot)^{10}$, and $(\cdot) \xrightarrow{\!\!\times (\cdot)\!} (\cdot)^2 \xrightarrow{\!\!\times (\cdot)^2\!\!\!} (\cdot)^4 \xrightarrow{\!\!\times (\cdot)^4\!\!\!\!} (\cdot)^8 \xrightarrow{\!\!\times (\cdot)^2\!\!\!\!} (\cdot)^{10}$ can also be used to obtain the nonlinearity $(\cdot)^{10}$ with four multiplications.} of four multiplications $(\cdot) \xrightarrow{\!\!\times (\cdot)\!} (\cdot)^2 \xrightarrow{\!\!\times (\cdot)\!} (\cdot)^3 \xrightarrow{\!\!\times (\cdot)^2\!\!\!\!} (\cdot)^5 \xrightarrow{\!\!\times (\cdot)^5\!\!\!\!} (\cdot)^{10}$. 

\subsection{Proposed Linearizer}
\label{subsec:Proposed-Linearizer-Interp}
\begin{figure}[t]
	\centering{}\raggedright \includegraphics[scale=0.6]{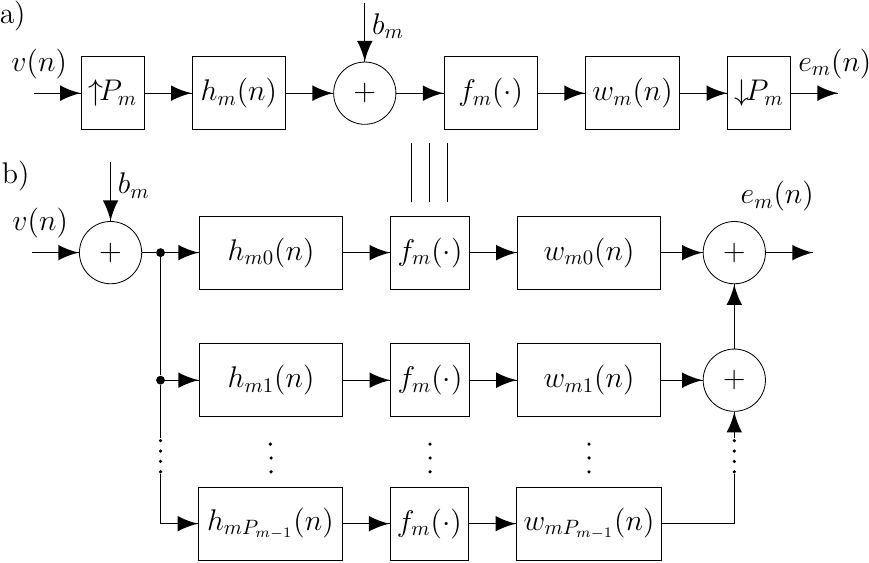}
	\caption{Nonlinear-branch implementation of the proposed linearizer.}
	\label{Flo:proposed-scheme_analog} 
\end{figure}

Utilizing again the polyphase decomposition of $H_k(z)$ in \eqref{eq:5} and that of $w_m(n)$ according to
\begin{equation}
	W_{m}(z)=\sum_{i=0}^{P_{m}-1}z^{i}W_{mi}(z^{P_{m}}),\label{eq:7}
\end{equation}
and the corresponding polyphase realizations, each nonlinear branch in the proposed linearizer can be implemented according to Fig. \ref{Flo:proposed-scheme_analog}(b), where the bias values have been moved to the input (see the explanation below). As seen, for the proposed linearizer, the static nonlinearities $(\cdot)^{p}$ present in the Hammerstein linearizer are again replaced with additive bias values followed by the simpler nonlinear modulus or ReLU operations (compare Figs. \ref{Flo:hammerstein-scheme_analog} and \ref{Flo:proposed-scheme_analog}).

\subsubsection{Implementation Complexity}\label{subsec:Proposed Linearizer-Interp-Complexity}
For the proposed linearizer with $N{=}K$, the number of multiplications is always \mbox{$(M{+}1)\times(K{+}1$)} as there are no multiplications involved in the nonlinear operations. If implemented straightforwardly, this comes at the cost of an increase in the number of bias additions as $P_m$ such additions are required in each nonlinear branch. However, observing that the polyphase components approximate unity-gain allpass filters \cite{Johansson_05}, one can carry out only one bias addition at the input of each nonlinear branch as seen in Fig. \ref{Flo:proposed-scheme_analog}. This is because a constant $b_m$ at the input of a linear and time-invariant filter $h_{mk}(n)$ results in the constant $b_mH_{mk}(e^{j0}){=}b_m$ at its output for a unity-gain allpass filter with a real impulse response $h_{mk}(n)$ which is the assumption here. In this way, the overall number of bias additions will be the same as in the proposed pre-sampling scheme in Section~\ref{sec:Proposed-Linearizer}. 

\subsection{Simulations and Results}
\label{subsec:Simulations-and-results2}
\begin{figure}[t]
\begin{centering}
\includegraphics[scale=1.00]{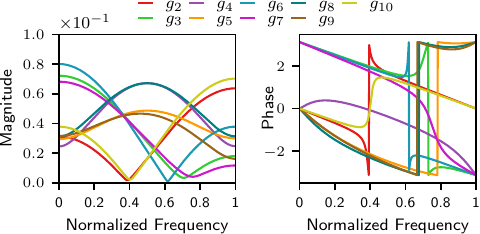} 
\par\end{centering}
\caption{Magnitude and phase responses of $g_{p}(k)$, for $p{=}2,\cdots,10$,
in \textit{Example 4}.}
\label{Flo:mag_phase2.1} 
\end{figure}

We assume a distorted signal $v(n)$ generated in the analog domain through the discrete-time equivalence in Fig. \ref{Flo:discrete-time-equivalence2}. The nonlinear-model coefficients were chosen in the same way as before (see Footnote 4), thus here with $g_{0}{=}0$, $g_{1}(k){=}[0,1,0]$, and $g_{p}(k)$ randomly generated for $k{=}0,1,\ldots,D$, $p{=}2,3,\ldots,Q$ with $D{=}2$ and $Q{=}10$.  We thus assume a distorted signal $v(n)$ exhibiting second-order filtered distortion, with $g_{p}(k)$ randomly generated with frequency responses as seen in Fig. \ref{Flo:mag_phase2.1}.

\begin{figure}[tbp]
\begin{centering}
\includegraphics[scale=1.15]{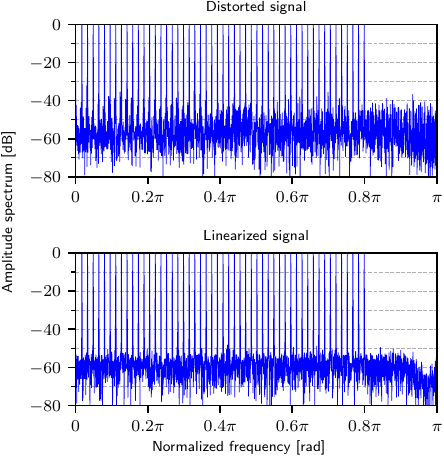} 
\par\end{centering}
\caption{Spectrum before and after linearization for a multi-sine signal using the proposed linearizer with a filter order of $M{=}6$ and $N{=}16$ nonlinear branches (\textit{Example 4}).}
\label{Flo:Spectrum4} 
\end{figure}

\begin{figure}[t]
\begin{centering}
\includegraphics[scale=0.98]{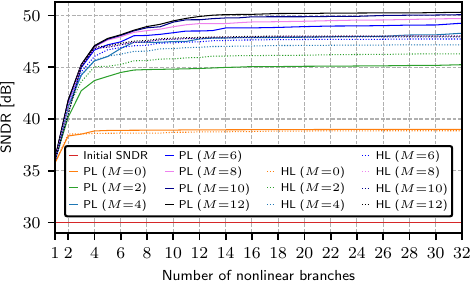} 
\par\end{centering}
\caption{SNDR versus number of nonlinearity branches. Here PL stands for the proposed linearizer and HL for the Hammerstein linearizer. (\textit{Example 4})}
\label{Flo:SNDRvsBranch2.1} 
\end{figure}

\textit{Example 4: } We consider a set of multi-tone signals similar to those in \textit{Example 1}, and \textit{Example 3}, generated through \eqref{eq:OFDM}, with $A_{k}{=}1$ for all $k$, $\omega_{k}$ computed as in \eqref{eq:omegak}, and $\alpha_{k}$ also randomly selected from $\{\pi/4,-\pi/4,3\pi/4,-3\pi/4\}$, corresponding to QPSK modulation. Here, we use $50$ active carriers out of $64$, covering approximately $80\%$ of the first Nyquist band. Both the reference and distorted signals are quantized to $10$ bits for a data set with $R{=}50$ and $R^{(eval)}{=}5000$ signals. Figure \ref{Flo:Spectrum4} displays the spectrum before and after linearization for one of the multi-tone signals, whereas Fig. \ref{Flo:SNDRvsBranch2.1} plots the mean SNDR over all signals for each instance of the proposed and Hammerstein linearizers.

In this example, the SNR is approximately 53 dB without distortion, whereas the SNDR is about 30 dB for the distorted signals before linearization. As seen in Fig. \ref{Flo:SNDRvsBranch2.1}, with a linearizer filter order of $M{=}2$, the SNDR can be enhanced by about $15$ dB. With $M{=}4$, an additional 3 dB SNDR improvement can be achieved. Increasing the order to $M{=}12$ can further enhance the SNDR by $3$ dB, totaling a $21$ dB improvement and approaching $51$ dB, which is about $2$ dB below the SNR. Beyond this level, additional increases in the linearizer filter order yield progressively smaller improvements, making further enhancements computationally inefficient. Hence, as for the pre-sampling linearizer, there is a clear trade-off between additional computational complexity and SNDR improvement. For the Hammerstein linearizer, the SNDR saturates at a somewhat lower level due to $L_{2}$-regularization, which is needed to keep the multiplier values small to avoid large quantization noise amplification, as discussed earlier in Section \ref{sec:Proposed-Linearizer}. {Finally, comparing this example with Examples 1-3, where 12-bit data was used, it is seen that the data wordlength does not affect the robustness, but only the SNDR levels (compare Figs. 7 and 21).}

\subsubsection{Implementation Complexity}
\label{subsec:complexity2} 

\begin{figure}[tbp]
\begin{centering}
	\includegraphics[scale=0.98]{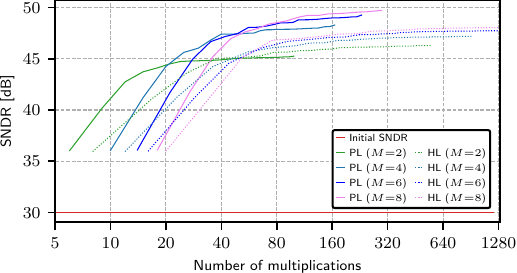} 
	\par\end{centering}
\caption{SNDR versus number of multiplications in\textit{ Example 4}. Here PL stands for the proposed linearizer and HL for the Hammerstein linearizer.}
\label{Flo:SNDR_versus_multiplications2.1} 
\end{figure}

An observation from Fig. \ref{Flo:SNDRvsBranch2.1} is that the Hammerstein linearizer appears slightly better than the proposed one when increasing the number of branches for the case with $M{=}2$. However, when comparing the SNDR against the number of multiplications for both methods, the proposed linearizer achieves a higher SNDR with lower complexity. This is illustrated in Fig. \ref{Flo:SNDR_versus_multiplications2.1}.

It is seen that for any chosen Hammerstein linearizer configuration, there is always a configuration in the proposed method that achieves superior performance with lower complexity. {The savings range from a few percent up to about 60\% depending on the scenario. For example, for $M{=}4$, to reach an SNDR of $46\,\text{dB}$, the proposed linearizer requires $30$ multiplications, whereas the Hammerstein linearizer requires $76$, corresponding to a saving of $46/76 {\approx}60.5\%$.}

Compared with the pre-sampling linearizers, the difference in complexity is larger here for the post-sampling linearizers. This is because the overall complexity of the static nonlinearities $(\cdot)^{p}$ is much higher here since several copies of them need to be implemented, and they have to be implemented separately. In the pre-sampling case, one can share computations between the different nonlinearities, which is why it suffices to compute $K$ multiplications in total to generate all nonlinearities $(\cdot)^{p}$ in Fig. \ref{Flo:Hammerstein-scheme}. Hence, in general, the proposed linearizer is more efficient in the post-sampling case than in the pre-sampling case, when comparing with the corresponding Hammerstein linearizers.

\begin{figure}[tbp]
	\begin{centering}
	\includegraphics[scale=1.15]{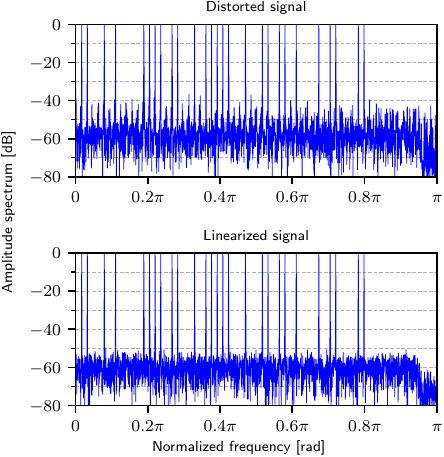} 
	\par\end{centering}
	\caption{Spectrum before and after linearization for a multi-sine signal with null subcarriers using the proposed linearizer with a filter order of $M{=}6$ and $N{=}16$ nonlinear branches (\textit{Example 5}).}
	\label{Flo:Spectrum5} 
	\end{figure}

\begin{figure}[tb]
\begin{centering}
\includegraphics[scale=1.15]{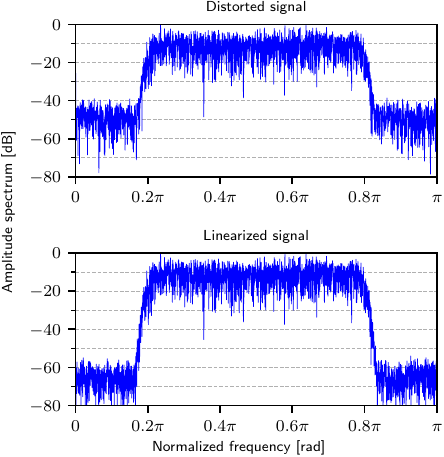} 
\par\end{centering}
\caption{Spectrum before and after linearization for a bandpass filtered white-noise signal using the proposed linearizer with a filter order of $M{=}6$ and $N{=}16$ nonlinear branches.} 
\label{Flo:Spectrum6} 
\end{figure}

\textit{Example 5: } To further illustrate the robustness of the proposed post-sampling linearizer, we have also evaluated it using the same type of multi-sine signal as in \textit{Example~4}, but with some subcarriers set to zero, and a bandpass filtered white-noise signal covering 60\% of the Nyquist band. As illustrated in Figs. \ref{Flo:Spectrum5} and \ref{Flo:Spectrum6} for each of these signals, essentially the same results are obtained as before. There is less than 1 dB SNDR degradation compared to the signals in \textit{\mbox{Example~4}} for which the linearizer was designed.

\section{Linearization Performance for Circuit-Simulated Data}
\label{sec:realdata}
\textit{Example 6:} {This example will demonstrate that the proposed design and linearizers also work when the reference signals are not known but have to be estimated. For this purpose, we assess the performance of the proposed linearizer applied on data obtained from circuit simulations in Cadence, and again compare with the Hammerstein linearizer.} 

We use a dataset of distorted single-tone complex signals provided in an industry-collaboration project in which an internal non-commercial ADC was designed. The signals were initially generated at an RF frequency of 3.6 GHz and subsequently demodulated to their corresponding complex baseband signals. As this paper focuses on real signals, we use the real parts of the signals in the evaluation. Further, in the design of the linearizers, it is here necessary to estimate the reference signals, as only the frequencies are known but not the exact gain and phase offsets of the signals. We have used the least-squares based estimation technique detailed in Chapter 1.6 of \cite{lowenborg2006mixed} for this purpose.

The signal frequencies are $f{=}f_s \times [73, 93, 113]/L$, where the sampling frequency is $f_s{=}10$ GS/s and the signal length is $L{=}8192$. It corresponds to the rounded frequencies $[89.1, 113.5, 137.9]$ MHz. The signal level is approximately $-1$ dB full scale (dBFS). In the project, the focus was to improve the spurious-free dynamic range (SFDR), measured in dBFS, which is why we use this metric here as well as the SNDR. Further, as the baseband signal bandwidth (below 138 MHz) is much smaller than the sampling frequency, we use the pre-sampling linearizers in Section \ref{sec:Proposed-Linearizer} (i.e., without additional interpolation).

It is observed that the signals contain high-order nonlinearity terms. The $x^{11}$-term is around $-65$ dBFS, whereas power terms above $11$ are below $-80$ dBFS. The nonlinearities also have a rather strong frequency dependency, which emanates from several sources (mixers, filters, and ADCs). This leads to relatively high filter orders for the linearizer filters when large SFDR improvements are targeted. This is because abrupt frequency dependent changes (corresponding to narrow transitions bands) require high filter orders \cite{Ichige_00}. For only a few signals, one may use lower filter orders, but then the quantization noise will be amplified due to ill-conditioned filter design. 

\begin{figure}[!t]  
	\begin{centering}
		\includegraphics[scale=1.15]{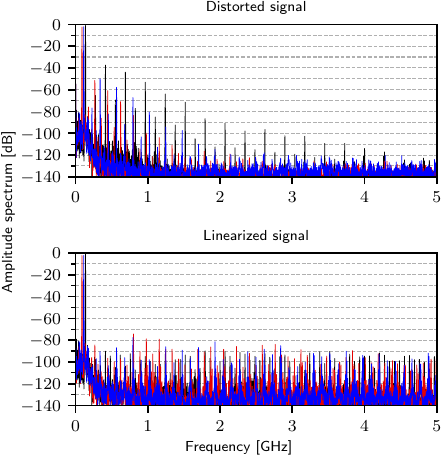}  
		\par\end{centering}
	\caption{Spectrum before and after linearization for single-tone signals (superimposed) using the proposed linearizer in \textit{Example 6}.}	
	\label{Flo:RD_spectrum}
\end{figure}

Using the proposed linearizer with $9$ branches, to suppress all the nonlinearities to around $-75$ dBFS we need a filter order of $M{=}22$. For this design, the SNDR is $68$ dB. The Hammerstein linearizer with the same number of branches and filter order achieves practically the same SFDR and SNDR. Recall that the performance of the proposed pre-sampling linearizer is comparable to that of the corresponding Hammerstein linearizer for larger values of $M$, both regarding SNDR (and SFDR) improvements and implementation complexity (see the discussion in Section \ref{subsec:complexity1}). Again, however, the proposed linearizer has the advantage that it does not require internal data quantizations. The spectra of the linearized signals are as shown in Fig. \ref{Flo:RD_spectrum} when using the proposed linearizer. 

{For comparison, commercial GS/s-rate 12-bit ADCs operating in similar bandwidth regimes typically report ENOB values around 8--10\, (corresponding to SNDR values around 50--62 dB) and SFDR values in the 65--78\,dBFS range under standard single-tone testing (see for example \cite{TI_datasheet}). Hence, the 68 dB SNDR and 75 dBFS SFDR obtained here after linearization represent state-of-the art performance.}

\section{Alternative Nonlinear Functions}\label{sec:alternative-nonlinearities} 
{In this paper, the nonlinear functions $f_{m}(v)$ are chosen as either the modulus or ReLU due to their simplicity and low complexity in hardware implementation \cite{Tarver_2019,deiro23}. We have also considered other common nonlinear functions such as the sigmoid, hyperbolic tangent, and exponential linear unit (ELU), but they significantly increase the implementation complexity for a targeted SNDR, as they require additional multiplications and additions. To illustrate this, Fig.~\ref{Flo:nonlinearities} shows, for the second-order distortion in \textit{Example~3}, the performance of the different nonlinear functions in terms of SNDR versus the number of multiplications. As seen, the use of these alternative nonlinear functions leads to worse performance and substantially increased complexity compared to the use of the modulus (or ReLU) as well as the Hammerstein nonlinearities. Here, the complexities when using the alternative functions are based on their Taylor expansions with three and five terms, and not counting trivial multiplications like 1/2. The plot in Fig.~\ref{Flo:nonlinearities} also includes the nonlinear function leaky ReLU (with negative-slope coefficient $\alpha {=} 0.1$) which leads to a linearizer with approximately the same complexity as for the Hammerstein linearizer (with the same number of branches), but also with worse linearization performance when the number of branches is relatively small.
\begin{figure}[!t]
			\centering
			\includegraphics[scale=0.98]{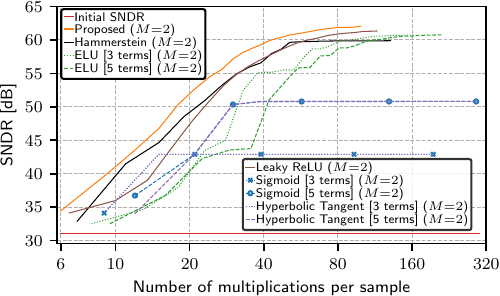}
			\caption{SNDR versus the number of multiplications for different nonlinear functions.}
			\label{Flo:nonlinearities}
\end{figure}}

{Another alternative is the simple binary step function which corresponds to 1-bit quantization. Thereby, the multiplications can be eliminated as their inputs are either zero or one. However, it was shown in~\cite{deihabit} for the frequency-independent case ($M{=}0$, the only case considered in that paper), that the linearizer then requires at least an order of magnitude more branches than the proposed linearizer (and also the Hammerstein linearizer), in order to achieve the same moderate SNDR level (eight-bit data). For higher SNDR levels, even two orders of magnitude more branches may be required. This implies that the number of additions required will increase ten times or more and implies that the total implementation complexity may even exceed that of the proposed linearizer. It was also shown in~\cite{deihabit} that, with properly selected bias values, an alternative implementation based on look-up tables can be used, thereby eliminating all computations except for one addition and one multiplication. The price to pay is the implementation cost of the look-up table whose memory size equals the number of branches plus one ($N{+}1$). Thus, that alternative is attractive primarily for the frequency-independent case and moderate resolutions. For the general frequency dependent case studied in this paper ($M{>}0$), an extension of that approach would become less attractive as the cost of the memory then becomes high, especially for higher resolutions requiring many branches and because the memory size is here $(N{+}1) {\times} (M{+}1)$. It is also noted that the look-up table approach proposed in~\cite{deihabit} can only be extended to the pre-sampling linearizer in this paper (Fig.~\ref{Flo:proposed-scheme1}), not the post-sampling linearizer (Fig.~\ref{Flo:proposed-scheme_analog}). For the latter, one may consider using a look-up table in each branch, but then the total memory size would grow exponentially as $(N{+}1) {\times} 2^{M{+}1}$ which becomes prohibitive when $M$ and $N$ increase.}
	
{To conclude, considering frequency dependent as well as pre-sampling-distortion and post-sampling-distortion models, the modulus and ReLU are generally the most efficient choices for low-complexity linearization, especially for higher resolutions.}

\section{Conclusion}\label{sec:conclusion} 
This paper introduced low-complexity linearizers for the suppression of nonlinear distortion in ADIs. Two different linearizers were considered, based on models where the nonlinearities are incurred after and before sampling, respectively (referred to as pre-sampling and post-sampling linearizers, respectively). The proposed linearizers are inspired by neural networks but have an order-of-magnitude lower implementation complexity compared to traditional neural-network-based linearizer schemes. The proposed linearizers can also outperform the traditional parallel Hammerstein linearizers even when the nonlinearities have been generated through a Hammerstein model. This was demonstrated through numerous design examples for various signal types and scenarios, including both simulated and circuit-simulated data. In general, the analysis and simulations show that the proposed linearizer is more efficient in the post-sampling case than in the pre-sampling case, when comparing with the corresponding Hammerstein linearizers. An additional advantage of the proposed linearizer, over the Hammerstein linearizer, is that it eliminates the need for internal data quantizations, thereby automatically avoiding noise amplification from the quantizations to the output.

A design procedure was also proposed in which the linearizer parameters are obtained through matrix inversion. Thereby, one can circumvent the costly, time-consuming, and time-unpredictable iterative nonconvex optimization that is traditionally adopted for neural network training. It also offers predictable online training and real-time updates in response to changes in circuitry. The proposed design effectively handles a wide range of wideband multi-tone signals and filtered white noise. Simulations demonstrated significant signal-to-noise-and-distortion ratio (SNDR) improvements of about 20--30 dB for both the simulated and circuit-simulated data.
 
Further, related to hardware implementation cost, the proposed linearizers can achieve the same SNDR as the benchmark Hammerstein linearizers  with a lower computational (arithmetic) complexity which correlates with the hardware implementation complexity. One can therefore conjecture that the proposed linearizers can offer more efficient hardware implementations than Hammerstein. It was however beyond the scope of this paper to investigate hardware implementations but is left for future work. The focus of the paper was to assess the fundamental properties of the proposed linearizers and show that they are computationally more efficient than the Hammerstein linearizers.

{Finally, it is noted that the proposed linearizers resemble the Hammerstein linearizers in that the filters appear after the nonlinear operations. Future work will also consider cases where filters appear before the nonlinear operations, as in Wiener and Wiener-Hammerstein linearizers (see the discussion in Section~\ref{subsec:Contribution}).}

\appendices{}
\section{Least-Squares Solution}
\renewcommand{\theequation}{A.\arabic{equation}}
\setcounter{equation}{0} 

\label{sec:LS_Solution}
The equation $\mathbf{A_r}\mathbf{w}{=}\mathbf{b_r}$ can be solved in the least-squares sense by finding $\mathbf{w}$ that satisfies $\mathbf{A_r}^\top\mathbf{A_r}\mathbf{w}{=}\mathbf{A_r}^\top\mathbf{b_r}$, which yields $\mathbf{w}{=}(\mathbf{A_r}^\top\mathbf{A_r})^{-1}\mathbf{A_r}^\top\mathbf{b_r}$ \cite{hastie2009elements}. For a set of such equations, 
\begin{equation}
	\left\{
	\begin{aligned}
		\mathbf{A}_{1}\mathbf{w} & =\mathbf{b}_{1},\\
		\mathbf{A}_{2}\mathbf{w} & =\mathbf{b}_{2},\\
		 & \vdots\\
		\mathbf{A}_{R}\mathbf{w} & =\mathbf{b}_{R},
	\end{aligned}
	\right.
\end{equation}
the problem can be rewritten as
\begin{equation}
	 \mathbf{A}_{\text{stack}}\mathbf{w}=\mathbf{b}_{\text{stack}},
\end{equation}
with
\begin{equation}
\mathbf{A}_{\text{stack}}=\begin{bmatrix}\mathbf{A}_{1}\\
\mathbf{A}_{2}\\
\vdots\\
\mathbf{A}_{R}
\end{bmatrix},\quad\mathbf{b}_{\text{stack}}=\begin{bmatrix}\mathbf{b}_{1}\\
\mathbf{b}_{2}\\
\vdots\\
\mathbf{b}_{R}
\end{bmatrix},
\end{equation}
which yields
\begin{align}
\mathbf{w} & =(\underbrace{\mathbf{A}_{\text{stack}}^{\top}\mathbf{A}_{\text{stack}}}_{\mathbf{A}})^{-1}\underbrace{\mathbf{A}_{\text{stack}}^{\top}\mathbf{b}_{\text{stack}}}_{\mathbf{b}}.
\end{align}
This can be equivalently written as
\begin{equation}
\mathbf{w}=\bigg(\sum_{r=1}^{R}\mathbf{A}_{r}^{\top}\mathbf{A}_{r}\bigg)^{-1}\sum_{r=1}^{R}\mathbf{A}_{r}^{\top}\mathbf{b}_{r}. \label{eq:A3}
\end{equation}
Including $L_{2}$-regularization (a.k.a. Tikhonov regularization), a diagonal matrix $\lambda\mathbf{I}$ with small diagonal entries $\lambda$ is added to the right-hand side of \eqref{eq:A3}. The resulting equation is then equivalent to \eqref{eq:w7}, with $\mathbf{A}$ and $\mathbf{b}$ in \eqref{eq:w8} and $\mathbf{A}_{r}$ and $\mathbf{b}_{r}$ defined in Section \ref{subsec:Design}.
\begin{small} \bibliographystyle{IEEEtran}
\bibliography{references/references_main}
\end{small} 
\vspace{-20 pt}
\begin{IEEEbiography}[{\includegraphics
	[width=1in,height=1.25in,clip, keepaspectratio]{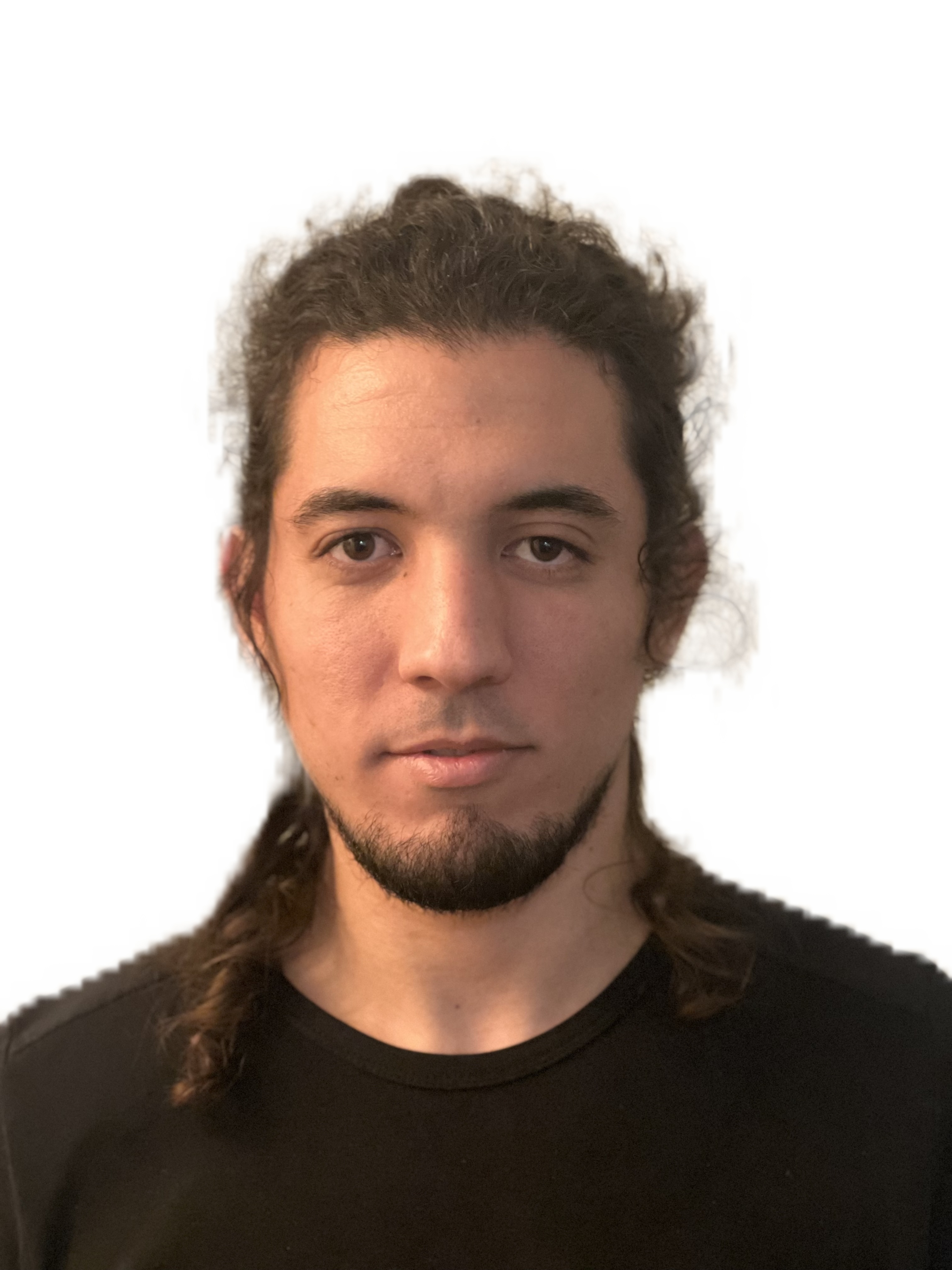}}]{Deijany Rodriguez Linares}
	(Graduate Student Member, IEEE) received the Bachelor of Science in Nuclear Engineering, a Postgraduate Diploma in Medical Physics and the Master of Science degree in Nuclear Engineering from the Higher Institute of Technologies and Applied Sciences (InSTEC), University of Havana, Cuba, in 2015, 2016 and 2018, respectively. He is currently pursuing a Ph.D. degree with the Division of Communication Systems, Department of Electrical Engineering, Link\"oping University, Sweden.
	From 2019 to 2020, he was an Associate Researcher at InSTEC, and from 2015 to 2018, he worked as a Junior Medical Physicist at the Cuban State Center for the Control of Drugs, Equipment and Medical Devices (CECMED). Since 2019, he has been a Junior Associate of the Abdus Salam International Centre for Theoretical Physics (ICTP), Trieste, Italy. His research interests include signal processing, wireless communication, reinforcement learning, and mathematical optimization.
\end{IEEEbiography}
\vfill
\pdfbookmark[1]{Biography: H\r{A}kan Johansson}{bio-hakan}
\begin{IEEEbiography}[{\includegraphics
	[width=1in,height=1.25in,clip,
	keepaspectratio]{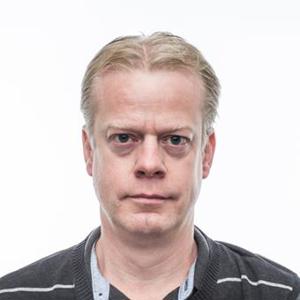}}]{H\AA kan Johansson}
	(S'97--M'98--SM'06) received the Master of Science degree in Computer Science and Engineering, and the Licentiate, Doctoral, and Docent degrees in Electronics Systems, from Link\"oping University, Sweden, in 1995, 1997, 1998, and 2001, respectively. During 1998 and 1999 he held a postdoctoral position with the Signal Processing Laboratory, Tampere University of Technology, Finland. He is currently a Professor at the Division of Communication Systems, Department of Electrical Engineering, Link\"oping University. He was one of the founders of the spin-off company Signal Processing Devices Sweden AB in 2004 (now Teledyne SP Devices). His research encompasses theory, design, and implementation of efficient and flexible signal processing systems for various purposes. He has authored or co-authored four books and some 80 journal papers and 150 conference papers. He has co-authored one journal paper and two conference papers that have received best paper awards and authored or co-authored three invited journal papers and four invited book chapters. He also holds eight patents. He served as a Technical Program Co-Chair for IEEE Int. Symposium on Circuits and Systems (ISCAS) 2017 and 2025. He has served as an Associate Editor for IEEE Trans. on Circuits and Systems I and II, IEEE Trans. Signal Processing, and IEEE Signal Processing Letters, and as an Area Editor for Digital Signal Processing (Elsevier).
\end{IEEEbiography}
\vfill\null

\EOD
\end{document}